\documentclass[sigconf,authorversion,nonacm,table]{acmart}
\AtBeginDocument{%
  }

\usepackage{soul}
\usepackage{url} 
\usepackage{subcaption}

\definecolor{purp}{RGB}{150, 59, 229}
\usepackage{lscape}
\usepackage{longtable}
\usepackage{multirow}
\newenvironment{mytabular}[1][1]{%
  \renewcommand*{\arraystretch}{#1}%
  \tabular%
}{%
  \endtabular
}

\colorlet{usercolorname}{yellow!0}
\sethlcolor{usercolorname}

\copyrightyear{2026}
\acmYear{2026}
\setcopyright{cc}
\setcctype{by-nc-nd}
\acmConference[CHI '26]{Proceedings of the 2026 CHI Conference on Human Factors in Computing Systems}{April 13--17, 2026}{Barcelona, Spain}
\acmBooktitle{Proceedings of the 2026 CHI Conference on Human Factors in Computing Systems (CHI '26), April 13--17, 2026, Barcelona, Spain}
\acmPrice{}
\acmDOI{10.1145/3772318.3790344}
\acmISBN{979-8-4007-2278-3/2026/04}

\begin{document}

\title[Operationalizing Perceptions of Agent Gender]{Operationalizing Perceptions of Agent Gender: \texorpdfstring{\\}{}  Foundations and Guidelines}


\author{Katie Seaborn}
\orcid{0000-0002-7812-9096}
\affiliation{%
  \department{Department of Industrial Engineering and Economics}
  \institution{Institute of Science Tokyo}
  \city{Tokyo}
  \country{Japan}
}
\affiliation{%
  \department{Department of Computer Science and Technology}
  \institution{University of Cambridge}
  \city{Cambridge}
  \country{UK}
}
\email{katie.seaborn@cst.cam.ac.uk}

\author{Madeleine Steeds}
\orcid{0000-0003-3767-292X}
\email{madeleine.steeds@ucd.ie}
\affiliation{%
  \institution{University College Dublin}
  \city{Dublin}
  \country{Ireland}
}


\author{Ilaria Torre}
\orcid{0000-0002-8601-1370}
\affiliation{%
  \institution{Chalmers University of Technology and University of Gothenburg}
  \city{Gothenburg}
  \country{Sweden}
}
\email{ilariat@chalmers.se}


\author{Martina De Cet}
\orcid{0009-0003-3968-3240}
\affiliation{
  \institution{Chalmers University of Technology and University of Gothenburg}
  \city{Gothenburg}
  \country{Sweden}
}
\email{demart@chalmers.se}

\author{Katie Winkle}
\orcid{0000-0002-3309-3552}
\affiliation{%
  \department{Department of Information Technology}
  \institution{Uppsala University}
  \city{Uppsala}
  \country{Sweden}
}
\email{katie.winkle@it.uu.se}

\author{Marcus G{\"o}ransson}
\orcid{0009-0009-9650-7917}
\affiliation{%
  \institution{Uppsala University}
  \city{Uppsala}
  \country{Sweden}
}
\email{marcus97goransson@gmail.com}

\renewcommand{\shortauthors}{Seaborn et al.}

\begin{abstract}
The ``gender'' of intelligent agents, virtual characters, social robots, and other agentic machines has emerged as a fundamental topic in studies of people's interactions with computers. Perceptions of agent gender can help explain user attitudes and behaviours---from preferences to toxicity to stereotyping---across a variety of systems and contexts of use. 
Yet, standards in capturing perceptions of agent gender do not exist.
A scoping review was conducted to clarify how agent gender has been operationalized---labelled, defined, and measured---as a perceptual variable. 
One-third of studies manipulated but did not measure agent gender. Norms in operationalizations remain obscure, limiting comprehension of results, congruity in measurement, and comparability for meta-analyses. The dominance of the gender binary model and latent anthropocentrism have placed arbitrary limits on knowledge generation and reified the status quo. We contribute a systematically-developed and theory-driven meta-level framework that offers operational clarity and practical guidance for greater rigour and inclusivity.
\end{abstract}





\maketitle

\section{Introduction}
\label{sec:intro}

\emph{Gender} (noun) is a complex factor of human identity and experience. While definitions vary and can conflict~\cite{Hyde2019}, general academic and social consensus unite on gender as ``how individuals and groups perceive and present themselves within specific cultures'' in terms of the ``psychological, social and cultural factors that shape attitudes, behaviours, stereotypes, technologies and knowledge''~\cite[p. 138]{Tannenbaum2019}. Gender touches all people and societies, albeit in different ways. Gender factors into how we self-identify~\cite{tajfel2004social,Reimer2022selfcat,Risman2018gender}, how we are identified and understood by others~\cite{Reimer2022selfcat,Alt2024,Tannenbaum2019,Risman2018gender}, and how we relate to each other~\cite{butler2013gender,Alt2024,Hyde2019,Allen2009power,Risman2018gender}. Gender organizes human life, dictating expectations for people classified as a certain gender: from appearance to social roles to abilities to occupations~\cite{butler2013gender,Butler2020,Hyde2019,Alt2024,Tannenbaum2019,Reimer2022selfcat,Risman2018gender}. Power is shaped by gender, with men and masculinity typically at the top and sexism against women and gender minorities continuing to be a global issue~\cite{Seaborn2023transce,Jewkes2015,Allen2009power,Risman2018gender}. Even while varying by culture and time, gender is among the most persistent social constructs in human history~\cite{Risman2018gender,Steinhagen2025reconstruct}.


The human tendency \emph{to gender} (verb) non-human agents~\cite{perugia2023robot,Seaborn2022pepper} 
has led to a vast and vibrant body of work on \emph{anthropomorphism}: the un/conscious application or interpretation of humanlikeness to non-human entities, living or machine~\cite{weiss2020inconsequential,bhatti_what_2023,spatola_human-robot_2019}. The fields of human-robot interaction (HRI) and human-agent interaction (HAI) evolved in part due to this feature of the design of social robots, intelligent virtual agents (IVAs), chatbots, and other interactive agents. Early and ongoing work on phenomena like the uncanny valley~\cite{mori_uncanny_2012} and the Computers are Social Actors (CASA) paradigm~\cite{nass_are_1997} has revealed that people tend to gender agents, even when only marginal cues to humanlikeness are present---and even when these feel ``off'' or uncanny---often without realizing it and despite knowing that the agent is a machine~\cite{nass_machines_2000,nass_are_1997,nass_wired_2005}. We cannot help but flex our mental models of human gender and other identities~\cite{seaborn2025socialidentity} when our brains are processing stimuli that bear resemblance to our species. Gender, being a core feature of humanity, is a readily available and easily understood model.

Understanding agent gendering has thus become common within human-computer interaction (HCI) research~\cite{perugia2023robot,Seaborn2022pepper,DeCet2025ambig}. Voice or body~\cite{Sutton2020,DeCet2025ambig,song_mind_2020,seaborn_voice_2021}, role or occupation~\cite{parlangeli2023stereotypes,galatolo2023right, jackson_exploring_2020,torre2023canagender,rogers2020robotgendering, miranda2024case}, mannerisms and behavioural responses~\cite{urakami_nonverbal_2023,kraus2018effects,borau_most_2021,winkle2021boosting}---various gender factors are now acknowledged as design features of anthropomorphic interactive agents.
Here, we raise a critical gap in practice
: the absence of a robust, meta-level approach to understanding \emph{whether} and \emph{how} people gender agents. Driving this state of affairs are two issues. First, \emph{operationalizations} of agent gender perceptions are unclear or missing in reporting. Succinctly, an operationalization comprises (i) an abstract \emph{concept} with (ii) a \emph{conceptual definition} grounded in \emph{theory} and a clear \emph{epistemological stance}, one that covers all conceptual \emph{dimensions} and links to (iii) one or more measurable \emph{indicators} or \emph{measures} and (iv) their \emph{measurement}, each capturing a conceptual dimension~\cite{sage2017oper,sage2022oper}. The second issue relates to operational follow-through: missing \emph{standards} for measuring perceptions of agent gender. Fundamentally, this is a problem of transparency, standardization, and potential epistemological conflation in practice and reporting. While an important body of work has sought to guide collection and reporting standards for \emph{participant} gender~\cite{Burtscher2020,schlesinger2017intersectional, spiel_patching_2019,jaroszewski_genderfluid_2018,offenwanger_diagnosing_2021,spiel_how_2019,winkle202315}, i.e., real humans, the same has not been done for non-human \emph{agents}. 
Moreover, while implicit tools exist, like stereotype models based on human gender norms~\cite{oliveira2019stereotype}, such approaches are premised on \emph{human} models that may not transfer or address all aspects of agent gendering, and thus do not directly reveal what, if any, gender percepts are at play. Finally, the scant evidence available indicates that experts may prescribe or limit gender response options~\cite{Seaborn2022pepper,seaborn_unboxing_2025}, thus distorting our understanding of whether and how people gender agents. We seem to need a standard meta-level approach to operationalization and a theory-inclusive, validated set of methods for directly measuring perceptions of agent gender.

To this end, we scoped out the state of affairs in pursuit of developing a theory-driven and consensus-based meta-level framework of terms, theories, and measures. Our main research question (RQ) was: \textbf{\emph{How should agent gendering be approached by the communities of practice at a meta-level?}}
First, we asked
\emph{RQ1: How have perceptions of agent gender been approached in human-agent interaction work so far?}
This set a foundation 
via the current scope of operationalizations, comprising terms and concepts, theory, and methodology, notably measurement. Second, we asked: 
\emph{RQ2: What consensus is there on how perceptions of agent gender can be operationalized and measured in a standardized way?}
Our findings reveal great diversity and an absence of standards and theoretical foundations in the literature. In response, we propose a novel meta-level framework and offer a launching pad for lower-level operationalization development grounded in the surveyed work and informed by theory and practice within and beyond studies of human-agent interactions. Our work establishes a baseline meta-level standard alongside a guide for best practice and next steps, with the caveat of future refactoring given the dynamic nature of gender and practice.

\section{Theory and Axioms}
We first outline our theoretical framework and axioms: the assumptions underlying our philosophy and epistemological approach. 

\subsection{Gender and Gendering}
We begin by recognizing that gender is defined in a variety of ways that may be complementary, conflicting, and/or contested~\cite{Hyde2019}. We offer an outline of the concept with respect to extant work. 

We define gender as a \emph{social construct} embedded in people's mental models of the human world. Social constructivism is a meaning-making process: societies and cultures agree and rely upon shared meanings attached to various entities, actions, and events~\cite{burr2015social}. 
Most societies have had the notions of \emph{male, men, and masculinity} alongside \emph{female, women, and femininity}, plus other genders, even if small in number and marginalized. Yet, because gender is constructed, the social response and even the meaning of gender can shift and vary over time and across individuals. For example, genderqueer and non-binary identities have gained social traction, even while thought to be held by few people~\cite{richards2016non}. \emph{Gender variance} is a staple of humanity~\cite{paramo2024ending}, even if dominant modes of gender exist.

We also recognize gender as \emph{intersectional}. Intersectionality is a legal and social framework that shows how power intersects among social characteristics---gender but also race, age, disability, and more---such that the experiences, privileges, and disadvantages at a given intersection are irreducible~\cite{crenshaw2013mapping,collins2020intersectionality}. 
For instance, social justice and intersectional work has revealed a legacy of colonial suppression and elimination of gender identities within Indigenous communities, like Two-Spirit~\cite{Davis2019}. 
``Two-Spirit'' is intersectional because it relates to gender and also biological sex, sexual orientation, and other social and cultural factors beyond the Western-centric ``LGBTQI+'' model~\cite{Davis2019}. Moreover, those with Two-Spirit identities experience different and more severe inequities and discrimination compared to, for instance, women (gender) and queer people (sexual orientation)~\cite{kia2020poverty}.
The colonial institution of Canada has meted out physical and cultural genocide against the Indigenous nations of Turtle Island~\cite{Davis2019,neu2000accounting}. Still, as an institution driven by people, Canada was the first nation to recognize Two-Spirit 
identities in formal documents, like the census~\cite{statcan2022census}. Canada's mental model of gender has expanded in response to social demands.

We also recognize that gender is something that we actively \emph{do}~\cite{Butler2020}. This activity has implications for all those designing and implementing agents. We experts make decisions about whether and how agents leverage these acts of gender in their embodiment with users. As Butler explains, ``gender is instituted through the stylization of the body ... bodily gestures, movements, and enactments of various kinds [that] constitute the illusion of an abiding gendered self''~\cite[p. 519]{Butler2020}. 
The same applies to the automatons that we make in our own image(s). We constitute the ``makeup'' (in both senses) of the agents that we design, and when we manifest these agents as humanlike, gender enters the canvas.

Gender, conceptualized as a social construct that intersects with other factors of identity and power, then relates to three broader, intersecting variables: norms, identity, and relations~\cite{Hyde2019}:

\begin{itemize}
    \item \emph{Norms} are the hidden rules that guide our behaviour, learned as we participate in a given culture over time.
    \item \emph{Identity} refers to the internal and external representations we have of ourselves in the world. Identity is \emph{shaped by} and \emph{shapes} our interactions in a given socio-cultural context. For gender, \emph{gender roles} dictate identity conformity in a given society. For example, men are the breadwinners and women are the caretakers. 
    \item \emph{Relations}, or how we relate formally and informally, are guided by societal norms and our understanding of our own and others' identities.
\end{itemize}

Our understanding of gender norms, identities, and relations---conscious or not---guides our practice. We may assume, for instance, that voice assistants in servile roles should be voiced by women~\cite{bergen2016d}, designing such agents to accept verbal abuse and respond in a silly fashion because we are not women ourselves and do not have the lived experience of sexism. Thus, we reproduce negative stereotypes about and harmful relations with \emph{human} women~\cite{bergen2016d,Wang2020,winkle2021boosting}. We may be embarrassed when such problems are raised post mortem. To prevent this, we experts need to understand gender through self-education, engaging in reflexivity~\cite{Rode2011,CarstensenEgwuom2014}, and using ethical and power-sensitive approaches in practice~\cite{winkle2023feminist}.

We can translate the idea that, among people, gender is socially constructed to agent perceptions through the notion of \emph{user-centred design}. Agent gendering is an activity with undercurrents of power. We experts are in a position to offer (if not impose) the agent on the user. We should focus on meeting the needs, desires, and mental models of the user, while recognizing that agent designs may not be perceived as intended~\cite{Gulliksen2003}. 
This maps onto human--human theories of identity, including gender. \citet{butler2013gender} connected Foucault's notion of \emph{subjectification}~\cite{Foucault1982} to gender, whereby gendered subjects (i.e., agents) are made in a mutually vulnerable process between ourselves and others~\cite{Davies2020}. Subjects only become their gendered selves and constitute the gender of others by participating in self-other subjectification. Notably, agents cannot (yet) participate in subjectification when it comes to gender or identities~\cite{seaborn2025socialidentity}.
This notion also reflects the social identity approach (SIA)~\cite{hogg2016social,tajfel1979integrative,tajfel2004social,seaborn2025socialidentity}. 
We participate in a social process of \emph{identification} and \emph{identity creation}, even unconsciously: social identities, like gender, are a \emph{matter of perception} that rests on the perceiver's mental models of the world. Percepts may be shared by others (the designer, the researcher, other users) or not. 
They may be static or change over time. While a certain portion of people may perceive the gender of a given subject (the agent, here) in a certain, shared way, not everyone will: true generalization may not be possible. Any operationalization of gender perceptions will need to account for this.

\subsection{Perception and Percepts}
A \emph{perception} results from our interpretation of the sensory information from our sensory organs when processed by our brains~\cite{Schacter2016}. Perceptions are linked to stimuli in the real world~\cite{brunswik2023perception}. Yet, they are also influenced by our memories of the same or similar stimuli, our mental models, our biases, and other mechanisms of the brain that make up conscious and unconscious human cognition~\cite{brunswik2023perception,Schacter2016}. As noted by~\citet{seaborn2025socialidentity}, perceptions, including of non-human agents, are socially situated: interpreted through our understanding of the social environment, which is comprised of the social groups to which we belong and our own social and personal identities. This determines what kinds of perceptions emerge for individuals and groups and their nature. On gender, one social group at one time in history may have a binary set of options linked to social and occupational roles, while another group at another time might have multiple genders that may change over the lifespan or as a result of certain events, like power transference. Notably, we experts may be limited in what categories of gender we expect or acknowledge, which may differ from our participants~\cite{Seaborn2022pepper,seaborn_exploring_2022,buolamwini_gender_2018,Seaborn2023transce,seaborn_unboxing_2025} and influence our results in biased ways~\cite{Seaborn2022pepper,buolamwini_gender_2018,Seaborn2023transce}. Within the context of academic research, we capture user perceptions as indicators, measures, or dependent variables (DVs). 

Gender \emph{cues} are (un)consciously processed by our brains as perceptions~\cite{Alt2024}. 
This socially-situated process is driven by internalized models of gender and individual experience~\cite{Alt2024}, with likely transference to non-human agents~\cite{seaborn2025socialidentity,nass_are_1997}. 
Numerous cues can elicit gender perceptions. In embodied agents, appearance is key~\cite{seaborn2022neither, Seaborn2022pepper}: facial features~\cite{ghazali2018effects}, shape~\cite{perugia2023robot}, makeup~\cite{acskin2023gendered}, and clothing~\cite{jung2016feminizing, perugia2023robot}. Auditory cues can signal gender, like voice pitch~\cite{fujii_intersectional_2025, seaborn_voice_2021}. Self-expressive linguistic cues like pronouns~\cite{fujii_intersectional_2025, fujii_silver-tongued_2024} or physical movement~\cite{zibrek15exploring} carry gender connotations. Cues are shaped by the context and the individual, whether it be social knowledge (like gendered names~\cite{feine2019gender}) or social situatedness (like occupational stereotypes~\cite{wang2021analysis} and gender norms~\cite{acskin2023gendered}).
These factors are also culturally sensitive. For example, Japanese gendered first-person pronouns and self-referents translate to agents~\cite{fujii_silver-tongued_2024,fujii_intersectional_2025}. Creators may have differing mental models and perceptions of gender across cultural contexts. The ambiguously-designed social robot Pepper, for example, with its body shape and country-specific gendered marketing, may be perceived as a ``boy'' in Japan and a ``woman'' in the West~\cite{Seaborn2022pepper}. Moreover, gender ambiguity can result in unexpected gendering across individuals and populations~\cite{Seaborn2022pepper}. Measuring user perceptions---rather than assuming the designed cues will be perceived as intended---is crucial for understanding reality.

\subsection{Agents and Agentic Computers}

The notion of \emph{agent} has long held a central place in HCI. ``Agent'' broadly refers to systems that act on behalf of users or exhibit goal-directed behaviour~\cite{wooldridge1995intelligent}. Notably, agents do not necessarily interact with human users. In early work, agents were often envisioned as autonomous or semi-autonomous software entities, like the ``interface agents'' from the 1990s~\cite{lashkari1994collaborative} designed to assist users by learning preferences and taking initiative. Software agents were intended to automate repetitive, poorly specified, or complex processes by bridging the ``gulfs between a user's desires and the actions that could satisfy them''~\cite[p. 68]{lewis1998designing}. The term has historically been difficult to define, with early work even proposing a taxonomy of agents based on biological models, accounting for kingdom, phylum, class, etc.~\cite{franklin1996agent}.

Over time, the notion of ``agent'' has expanded to include terms like intelligent virtual agents (IVAs), embodied conversational agents (ECAs), social robots, virtual humans, avatars, or socially interactive agents, among others~\cite{lugrin2022handbook}. In contemporary HCI and HAI, agency is often understood not solely in terms of autonomy or intelligent behaviour, but as a relational property. Drawing from a definition by the \href{https://hai-conference.net/what-is-hai/}{Human-Agent Interaction Conference}, we define an object or system as an agent when people interact with it \emph{as if} it had its own goals, intentions, and motivations. This is often captured under the CASA paradigm~\cite{nass_are_1997}. This perception is shaped not only by a system's capabilities, but also by its design---whereby designed social cues like anthropomorphism, speech, gestures, or personality come into play. Importantly, even systems with minimal autonomy may be treated as agents if people interpret their behaviour as intentional, while highly autonomous systems may not be perceived as agents at all if they lack interactive or social presence (e.g., a piece of technology like a telephone). In this view, agency emerges at the intersection of technological design and human interpretation, marking the core of HAI experiences.

\subsection{Unifying Theoretical Model: Gender as a Social Structure}
\label{sec:theorymodel}

A theoretical model that captures the human side (and potentially the human--agent realm) at all levels---from individual perceptions to societal influences---is \emph{gender as a social structure} by \citet{Risman2018gender,Risman2018structure}. This model maps out how the noun and verb of gender intersect within the larger social world, integrating and complementing other social constructivist theories like that of \citet{butler2013gender} for gender,  \citet{collins2020intersectionality} and \citet{crenshaw2013mapping} for intersectionality, and \citet{tajfel2004social} and \citet{turner_rediscovering_1987} for social identity. With its multi-level composition, the model acknowledges hierarchies and explains the role of gender from the individual to the group. 
The first level is the \emph{individual}, who---whether researcher or researched---may be passively socialized and actively participate in self-identity work, thereby developing general and unique percepts about concepts like agender gender. The second is \emph{interactional cultural expectations}, which occur among individuals in the social arena and involve other-categorization (and othering), biases, and power relations, like between researcher and participant. Encompassing both is the \emph{macro domain}, where institutions---including the law, dominant scientific standards, and ideological doctrines---hold influence.

The present work was motivated by the unrecognized gap in how researchers (as \emph{individuals}) have approached the operationalization and measurement of agent gendering, if at all, including whatever models and standards  (\emph{macro--institutional}) alongside biases and choices of measures and measurement  (\emph{interactional}) have been used in HAI research. Our systematic scoping review methodology aimed to surface this knowledge, as well as any gaps, to be filled with existing exemplars, should they exist, extant (and as yet unused) theory, and insights from other literatures.

\section{Methods}

We conducted a systematic scoping review~\cite{Munn2018}, following the PRISMA Extension for Scoping Reviews (PRISMA-ScR) protocol~\cite{PRISMA_checklist}, with minor modifications for our field of study. For example, we did not use a structured abstract or include ``scoping review'' in the title of this paper. Scoping reviews aim to broadly capture the state of affairs and generate a synthesis of the body of work so far~\cite{Munn2018}. Scoping reviews are ideal for new and active research topics and when meta-analysis may not be possible~\cite{Peters2015}. Peters et al. \cite[p. 141]{Peters2015} describe scoping review work as ``reconnaissance,'' where the goal is to identify and determine the extent to which consensus and conceptual clarity has been reached on terms and concepts. This makes the method ideal for a topic like operationalizing gender perceptions. Additionally, we use the systematic literature review procedure offered by the PRISMA-ScR~\cite{PRISMA_checklist} in pursuit of methodological standardization and rigour. The PRISMA-ScR~\cite{PRISMA_checklist} also provides guidance on how to draft up the manuscript, which we have followed, aside from the noted exceptions for our discipline. Our checklist is found in \autoref{appendix:checklist}.

\subsection{Protocol and Registration}
Our protocol was registered before data collection on June 1\textsuperscript{st} 2024 via OSF at \url{https://osf.io/t9m5c}. 

\subsection{Eligibility Criteria}
Inclusion criteria were: HAI and adjacent literature; and work in which gender perceptions of agents were measured or captured. We excluded items for the following reasons: inaccessible; inappropriate type, notably grey literature, literature reviews, unpublished reports, preprints, non-peer reviewed papers, conference papers and talks without proceedings (given our aim of using vetted and established approaches); and not written in a language known by authors (i.e., English, Japanese, Swedish, Italian). We also scoped the review to a 10 year period for manageability; this is common practice for scoping reviews (72\% in 2016)~\cite{Tricco2016scoping}.

\subsection{Information Sources}
We started with an existing dataset from a systematic review on 
the ACM/IEEE International Conference on Human-Robot Interaction (ACM/IEEE HRI), which covered years 2006--2022~\cite{Seaborn2023weird}. This work critically examined the demographic representativeness of participants in HRI research. The dataset acted as a quick-start baseline.
We used the search function in Google Sheets 
to filter based on our targeted 10 year timeline (2014--2024) and gender focus (using the keywords in \autoref{sec:search}).
We then conducted new searches of the ACM Digital Library (ACM DL) for coverage of the topical area (HAI broadly) and the 10-year timeframe. We also used Web of Science for a more general reach that included journal articles.

\subsection{Search}
\label{sec:search}
Aiming to include the items most relevant to the RQs, our query used keywords representing our agent and gender foci:

\begin{quote}
    \emph{(human-computer interaction* or human-robot interaction* or human-agent interaction*)} \textbf{(agent focus)}\\
    \emph{and}\\
    \emph{(gender* or sex)} \textbf{(gender focus)}
\end{quote}

We targeted the topic, title, abstract, and keyword metadata. We used filters to target the timeframe of 2014-2024 and exclude items based on article type, e.g., no systematic reviews.

\subsection{Selection of Sources of Evidence}
We had two screening phases. First,  we screened each item based on the title, abstract, and keyword/topics. Then, we checked the full text while making extractions, excluding items when necessary.

\subsection{Data Charting Process}
Screening and extraction was conducted independently in a Google Sheet. Given the number of items, the work was divided among all researchers. Each researcher prompted the team over email to double-check their work. In the screening phase, the venue and authors were hidden to reduce selection bias. We aimed to include rather than exclude, to err on the side of caution and fairness. If a researcher was uncertain about a particular item, they raised it with the team over email and another member double-checked it. 
In the extraction phase, the search function was used in the full text to find gender-related keywords by word stem (full term examples are placed in parentheses): (a/cis/trans/bi/etc.)gender, (inter)sex, (fe)male, (wo)men, (wo)man, masculine, feminine, (non)binary. The extractor also skimmed the theory, methods, and results sections, pertinent to the RQs. 
If an item was deemed inappropriate, it was marked and discussed among the team over email.

\subsection{Data Items}
Data items were: Metadata, including Paper ID (PID),
Venue,
Year,
Author(s),
Title,
URL,
Abstract,
Keywords;
Operationalizations, including
Term,
Definition,
Theory, and
Citation(s); and
Study Details, including
Study Goal,
Research Design,
Measure or Dependent Variable (DV),
Data Collection Method,
Measurement,
Subjective or Objective,
Number of Items,
Response Format,
Created/Existing,
Validated,
Self-Validation,
Validation Score,
Cronbach's Alpha,
Reference(s),
Qualitative Method, and
Qualitative Data Sources.

\subsection{Synthesis of Results}
We generated counts, percentages, and, where possible, quantitative statistics, like Cronbach's alpha averages
.
For the qualitative data, we carried out reflexive thematic analyses~\cite{Braun2006them}. We chose this approach in recognition of our situatedness as researchers within HAI and also gender studies, aiming to leverage our positionality while acknowledging and grappling with our personal frames towards the topic~\cite{Braun2006them}. Given the amount of data, we divided the work among us, e.g., one the measures, another the theory. We treated solo analysis as potentially bias-inducing, so we maintained vigilance about our own frames of reference and sought alternative perspectives from team members over email. Since our aim was to establish what was explicitly reported, our orientation was more semantic and realist; still, we took a critical and inductive frame to identify gaps and discrepancies. Each researcher familiarized themselves with the data under their care, generating codes and then categorizing these into initial higher-level themes. Each then went through a cyclical process of refining, naming, and defining the themes, keeping the RQs in mind. The initial and subsequent revisions of codes and themes were checked asynchronously by at least one other researcher when pinged over email; there was no secondary coding. Questions were raised over email, but there were no major issues or additions, simply requests for clarification.
The codebook is included in the open dataset (\url{https://bit.ly/genopoa}).

\subsection{Positionality}

We describe how our unique perspectives and relation to gender shaped our analytical processes, as a team and as individuals, in support of research credibility~\cite{Bardzell2011,hampton2021positionality}.
The first author treats gender as a social construct and a means of social categorization~\cite{seaborn2025socialidentity}. They identify as gender-diverse and gender-apathetic, which they believe allowed them to capture a broader pattern of gender concepts during data analysis. The second author is genderfluid and sees gender as diverse and changeable. In their analysis, they were open to broad categorisations of gender and concepts related to gender. 
Author three is a cisgender woman who grew up in a liberal environment within a conservative (by current standards) country; she has lived in several countries with different cultural views on gender, and has been involved in research and community initiatives aimed at reducing gender biases in society. She considers gender as a socially-constructed and contextual lived experience.
The fourth author lived most of her life in a conservative country, before moving to a more liberal one. This prompted reflection on gender as a social construct. 
Her analysis was influenced by being part of a queer community and her focus on gender diversity.
Author five considers gender a social construct and was informed by an intersectional, feminist approach to studying gender in the context of (norm-breaking) human-agent interactions. 
The last author considers gender a social construct. Before, he had studied gender to a limited extent in the context of HAI.

\section{Results and Findings}

\begin{figure*}[!ht]
  \includegraphics[width=.7\textwidth]{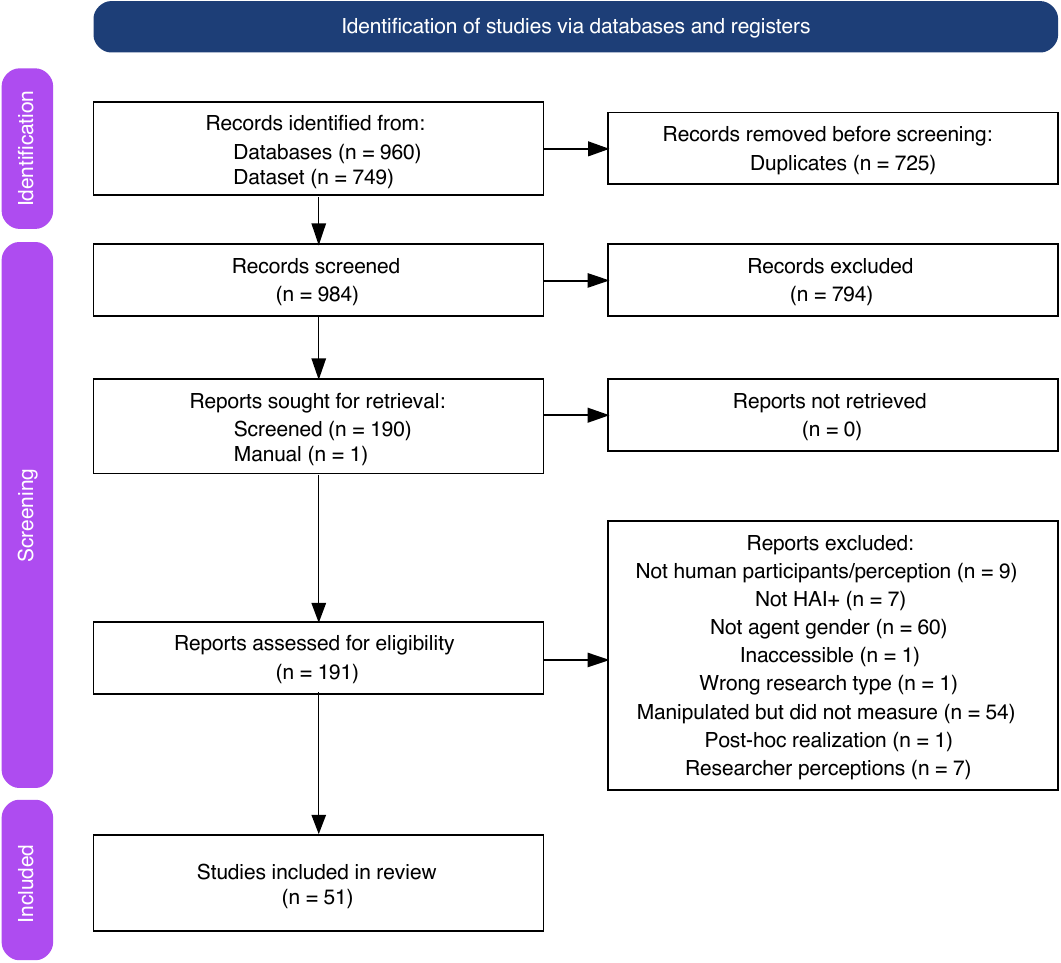}
  \caption{PRISMA flow diagram. Generated using the online tool by \citet{Haddaway2022prisma}.}
  \Description{PRISMA flow diagram showing the starting data set combined with a search of three databases resulting in 984 records screened and 191 records assessed, ultimately resulting in 51 records included.}
  \label{fig:prisma}
\end{figure*}

Of the 749 papers from \citet{Seaborn2023weird} and the 960 papers newly gathered, a final 51 were selected as eligible and included for data analysis (\autoref{fig:prisma}). Notably of the 191 assessed for eligibility, 54 excluded papers manipulated but did not measure gender (28.3\%).
Included papers were mostly published in CHI ($n=10)$, IJSR ($n=10)$, and HRI ($n=7)$. 

The open dataset is available at \url{https://bit.ly/genopoa}.

\subsection{Study Goals}
\begin{figure*}[!ht]
  \includegraphics[width=\textwidth]{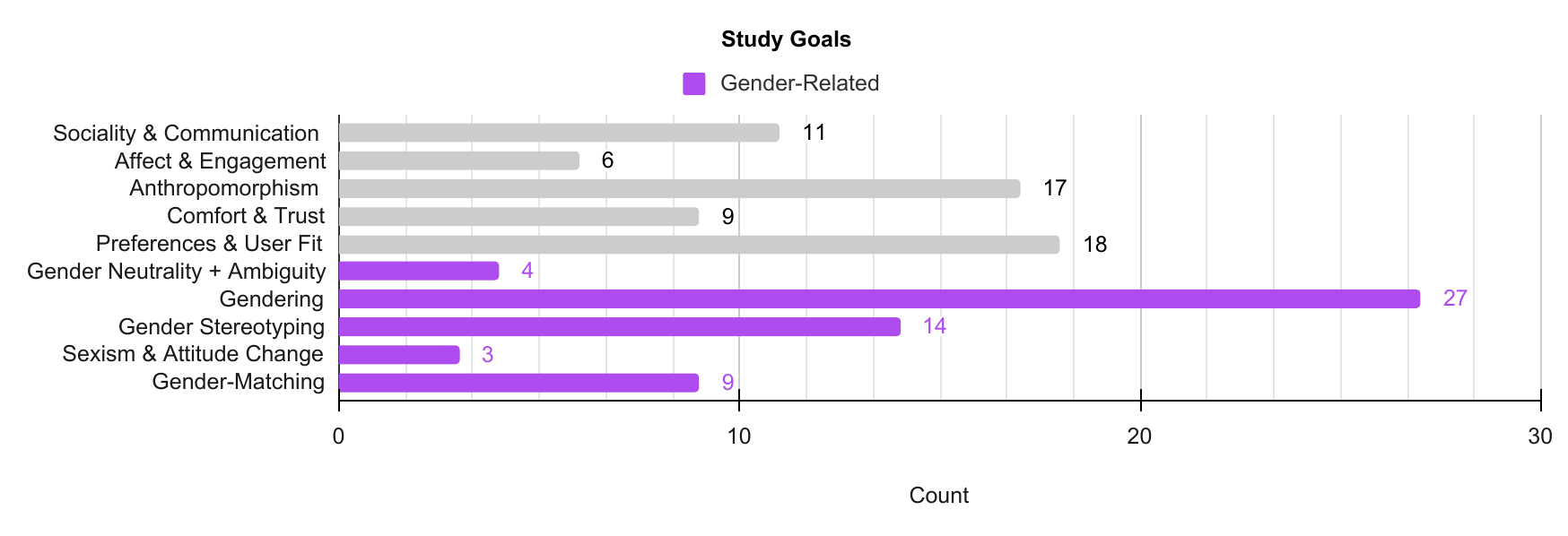}
  \caption{Study goals, with study counts by topic. Note: Each study could have multiple goals, so the sum does not equal $N=51$.}
  \Description{Bar chart indicating the counts of studies that are captured within each study goal topic. The most common topic was agent gendering, followed by anthropomorphism and preferences and user fit. The least common were sexism and attitude change and gender neutrality and ambiguity.}
  \label{fig:goals}
\end{figure*}

Study goals were gender-related or not (\autoref{fig:goals}). Most covered multiple topics (41, 79\%) beyond gender (15, 29\%). For example, \citet{rogers2020robotgendering} evaluated comfort levels and gendering of novel, neutrally-designed humanoid robots. \emph{Gender-related} topics included: 
\begin{itemize}
    \item Perceptions of agent gender neutrality and ambiguity (8\%), e.g., gendering intended gender-neutral robots~\cite{rogers2020robotgendering}
    \item Attitudinal and behavioural effects of agent gendering (53\%), e.g., trans and non-binary angles on a/gendered robots~\cite{stolp2024morethanbinary}
    \item Attitudinal and behavioural effects of agent gender stereotyping (27\%), e.g., agent voice pitch and propensity for task stereotyping~\cite{tolmeijer2021female}
    \item Anti-sexism and attitude change using gendered robots (6\%), e.g., uncovering sexist response design in South Korean voice assistant responses~\cite{Hwang2019sounds}
    \item Gender-matching of agent to user, role, or task (18\%), e.g., matching storyteller agent gender to child gender~\cite{robben2023theeffect}
\end{itemize}

Topics \emph{unrelated to gender} included:
\begin{itemize}
    \item Perceptions of agent sociality and communication (22\%), e.g., gaze differences between Japanese vs. Americans towards an agent~\cite{koda2018perception}
    \item Agent-related affective responses and engagement (12\%), e.g., an agent expressing anger~\cite{wessler2021economic}
    \item Perceptions of agent anthropomorphism (33\%), e.g., robot physicality in social situations~\cite{dennler2023design}
    \item Perceptions of comfort and trust for an agent (18\%), e.g., conversational agent design and trust responses~\cite{aljaroodi2023cultural}
    \item Preferences and user fit with an agent (35\%), e.g., Irish children's preferences for robot voices~\cite{sandygulova2015children}
\end{itemize}

Gender was also implicitly explored. For example, \citet[p. 52:2]{aljaroodi2023cultural} aimed ``to address ... how [cultural appropriateness] in avatar design affects users' trusting beliefs and usage intentions,'' implying a link between trust, intent, and agent gendering later explicated through the choice of measures and data analysis.

\subsection{Terms and Operationalizations}

\hl{682} gender-related terms were sourced.
\hl{Sixteen} examples of gender-related theorizing were found. This suggests assumptions of reader understanding and agreement on 
the term and underlying concept of ``gender.'' Yet, diversity and contradictions exist. The term and its operationalizations seem under-theorized.

Simple \emph{definitions of gender} included ``related to physical embodiment''~\cite{rogers2020robotgendering}, ``a social category''~\cite{reich2017irrelevance} or ``social stimuli''~\cite{huang2022proxemics}, and ``an important factor''~\cite{ye2020effect}. Others offered more in-depth and nuanced definitions and operationalizations. For example, \citet[pp. 984--985]{choi2020avatars} described gender as ``one of the most salient and commonly used variables that describe a person's characteristics'' ... ``operationalized with visual cues they imply a female or male avatar.'', while \citet[p. 2]{liu2023identity} distinguished sex and gender: ``differences between men and women through `sex' as a biological trait and ``gender' as a social and psychological identity.'' These surface the social identity dimensions of gender.

We now turn to \emph{theorizing and gender.} \citet[p. 1916]{acskin2023gendered} cited \citet{de1953second} and extant social models: ``gender is a \emph{formation}; it is something we \emph{do}~\cite{west1987} or something we \emph{realize}~\cite{butler2002gender}'' an ``active process referring to `doing gender' based on social relationships~\cite{de1953second}'' ... ``shaped according to cultural norms and stereotypes~\cite{butler2002gender}'' ... ``a non-compulsory category recurring according to norms~\cite{Morgenroth2018}.'' \citet[p. 3]{tolmeijer2021female} referenced theory-driving empirical work: ``Research on third gender associations has shown that typically people assign a gender to a voice, even though they cannot intuitively assign a gender~\cite{Sutton2020}.'' \citet[p. 449]{kuchenbrandt2012keepaneye} explained that gender ``is one of the most salient and omnipresent social categories in human societies that affects virtually every aspect in our every-day live'' ... ``gender determines people's social roles, occupations, relationships, and opportunities~\cite{Bussey1999}'' and ``the primary and most basic category of social perceptions of others and of the self~\cite{Harper2003}.'' \citet{fitton2023dancing} demarcated ``traditional'' from other gender definitions, but for reasons unknown: ``traditional associations between sex and gender (..) 
Participants were required to choose either male or female gender depending on which they identified with most.''. They cite ``A Guide To Gender Identity Terms'' on the National Public Radio (NPR) website\footnote{\url{https://www.npr.org/2021/06/02/996319297/gender-identity-pronouns-expression-guide-lgbtq}} without clear reason. \citet[p. 2]{pitardi2023congruity} define gender as ``a central dimension of individuals' self-concept and identity and is among the most important human characteristics that influence how people develop relationships with others (cf. \citet{Freimuth1982}).'' In short, gender definitions vary, can be vague (``important,'' ``social''), and can be contradictory (binary or not).

We have summarized the terms, concepts, and \emph{operationalizations} in a series of tables, linking in our theoretical model and the extant literature (\autoref{sec:theorymodel}) where appropriate. 
\begin{itemize}
    \item \autoref{tab:gendervocab} presents the \emph{gender vocabulary} used across the corpus of papers. This vocabulary may help the community speak with cohesion when reporting on agent gender work. 

    \item \autoref{tab:perceptions} offers gender perception processes, divided by actions, constructions, and judgments. These capture the ``how'' and ``what'' of agent gender perception work. The terms within each category may be interchangeable, with some nuances offered in the table. 
    \item \autoref{tab:varieties} showcases the range of genders and gender models used. We advocate for specifying use of specific models regardless of our own perspectives on gender, to distinguish fundamentally different operationalizations and measurements for meta-synthesis work. 
    \item \autoref{tab:power} offers concepts related to how power operates through gender at societal, cultural, and interpersonal levels. We highlight the power relationship noted in \autoref{sec:theorymodel} between the researcher and participant. 
    \item \autoref{tab:design} presents the methods used to elicit or control gendering through agent design. While only \citet{roesler2022context} mentioned gender manipulation as a concept, we emphasize that all design work---including choice of agent---may be an act of gender manipulation. 
    We suggest linking gender-related research goals to design manipulations when describing the agent and manipulation checks of agent gender.
\end{itemize}

\begin{table*}[ht]
\caption{Gender vocabulary.}
\label{tab:gendervocab}
\begin{tabular}{llp{185pt}p{190pt}}
\toprule
 & Vocabulary& Definition & Sources \\
 \midrule
\textit{Noun} & \textbf{Gender} & A social and cultural construct related to identity and relations; requires a specific operationalization. &  \autoref{sec:theorymodel}; 
all papers \\
\textit{Noun} & \textbf{Genderedness}  & The quality of being gendered (or not, with -less). & \cite{perugia2022bias,seaborn2023can,stolp2024morethanbinary,torre2023canagender,perugia2021gender,roesler2023hu,parlangeli2023stereotypes} \\
\textit{Adjective} & \textbf{Gendered} & The state of being gendered (or not, with non-). & \cite{huang2022proxemics,reich2017irrelevance,Telang2023,liu2023identity,roesler2022context,weiss2020inconsequential,seaborn2023can,tolmeijer2021female,Hwang2019sounds,moreira2023can,peck2020inducing,fitton2023dancing,stolp2024morethanbinary,chang2024investigating,otterbacher2017uncanny,sandygulova2018agegender,perugia2022bias,tay2014stereotypes,torre2023canagender,choi2020avatars,perugia2021gender,roesler2023hu,agren2022persuasiveness,thellman2018persuasive,bernotat2021female,parlangeli2023stereotypes,galatolo2023right,acskin2023gendered,nomura2023expectations,lee2021social,McGinn2020meet,stroessner2019perception} \\
\textit{Verb} & \textbf{Gendering} & The act of attributing gender, i.e., to gender. & \cite{galatolo2023right,rogers2020robotgendering,torre2023canagender,weiss2020inconsequential,moreira2023can,stolp2024morethanbinary,robben2023theeffect,pitardi2023congruity,roesler2023hu,acskin2023gendered,nomura2023expectations} \\
\textit{Adverb} & \textbf{Genderizing} & The act of attributing gender to a feature of the agent to invoke gendering, e.g., feminizing speech. & \cite{jung2016feminizing,roesler2022context,perugia2021gender,roesler2023hu,agren2022persuasiveness,McGinn2020meet,stroessner2019perception} \\
\bottomrule
\end{tabular}
\end{table*}

\begin{table*}[ht]
\caption{Gender perceptions.}
\label{tab:perceptions}
\begin{tabular}{lp{85pt}p{150pt}p{181pt}}
\toprule
Category & Processes & Definition & Sources \\
\midrule
\textit{Actions} & \textbf{Attributing Gender} & The process of classifying or assigning gender to an agent, conscious or not. & \cite{nomura2023expectations,robben2023theeffect,chang2024investigating,reich2017irrelevance,sandygulova2018agegender,perugia2022bias,torre2023canagender,roesler2022context,perugia2021gender,bernotat2021female,parlangeli2023stereotypes,acskin2023gendered,sandygulova2015children} \\
\textit{Constructions} & \textbf{Perceived Gender or Gender Perception} & The result of perceiving agent gender, i.e., the gender percept. & \cite{huang2022proxemics,perugia2022bias,roesler2023hu,shiomi2017audiovisual,Telang2023,sandygulova2015children,torre2023canagender,sandygulova2018agegender,jung2016feminizing,rogers2020robotgendering,deshmukh2018technology,seaborn2023can,stolp2024morethanbinary,tolmeijer2021female,chang2024investigating,otterbacher2017uncanny,tay2014stereotypes,kraus2018effects,roesler2022context,perugia2021gender,bernotat2021female,acskin2023gendered,McGinn2020meet,kervellec2016similarly,stroessner2019perception,dennler2023design} \\
 & \textbf{Gender Attribution} & The measured outcome of gendering an agent, i.e., the gender attribution. & \cite{weiss2020inconsequential,bernotat2021female,seaborn2023can,tolmeijer2021female,reich2017irrelevance,perugia2022bias,moreira2023can,ye2020effect,torre2023canagender,roesler2022context,sandygulova2018agegender,roesler2023hu,parlangeli2023stereotypes,acskin2023gendered,McGinn2020meet,tan2018isociobot,dennler2023design} \\
 & \textbf{Gender Distribution} & The range of attributed genders. & \cite{perugia2022bias,roesler2023hu,liu2023identity,aljaroodi2023cultural} \\
 & \textbf{Gender Schema} & The mental model of the gender percept. & \cite{acskin2023gendered,perugia2022bias,bernotat2021female,liu2023identity} \\
\textit{Judgments} & \textbf{Gender Congruity} & A (mis)match between agent and gender. & \cite{robben2023theeffect,pitardi2023congruity,reich2017irrelevance,otterbacher2017uncanny,tay2014stereotypes,torre2023canagender,kuchenbrandt2012keepaneye,roesler2022context,choi2020avatars,galatolo2023right,sandygulova2018agegender,roesler2023hu,thellman2018persuasive,wessler2021economic,bernotat2021female,acskin2023gendered,lee2021social,kervellec2016similarly} \\
 & \textbf{Gender-Normativity or Typicality} & Gendering, expected or assumed, based on socio-cultural norms and expectations. & \cite{reich2017irrelevance,kuchenbrandt2012keepaneye,stolp2024morethanbinary,sandygulova2018agegender,kraus2018effects,torre2023canagender,perugia2021gender,roesler2023hu,bernotat2021female,parlangeli2023stereotypes,galatolo2023right,acskin2023gendered,nomura2023expectations,aljaroodi2023cultural,stroessner2019perception} \\
 \emph{Reactions} & \textbf{Gender Effects} & Behavioural responses, reactions, and other outcomes of agent gender. & \cite{huang2022proxemics,rogers2020robotgendering,koda2017perception,tolmeijer2021female,moreira2023can,peck2020inducing,robben2023theeffect,chang2024investigating,otterbacher2017uncanny,reich2017irrelevance,ye2020effect,kraus2018effects,kuchenbrandt2012keepaneye,choi2020avatars,perugia2021gender,pitardi2023congruity,koda2018perception,sandygulova2018agegender,cameron2016congratulations,agren2022persuasiveness,shiomi2017audiovisual,thellman2018persuasive,wessler2021economic,liu2023identity,galatolo2023right,nomura2023expectations,lee2021social,aljaroodi2023cultural,kervellec2016similarly,stroessner2019perception} \\
 \bottomrule
\end{tabular}
\end{table*}

\begin{table*}[ht]
\caption{Gender varieties.}
\label{tab:varieties}
\begin{tabular}{lp{200pt}p{179pt}}
\toprule
Gender Varieties & Definition & Sources \\
\midrule
\textbf{Binary Gender} & The dominant two-factor gender model of man/woman, male/female, and masculine/feminine. & \autoref{sec:theorymodel}; most papers \\
\textbf{Traditional Gender} & Exclusively the gender binary; not always ``traditional'' across socio-cultural contexts and times. & \cite{fitton2023dancing,stolp2024morethanbinary,chang2024investigating,tay2014stereotypes,roesler2022context,choi2020avatars,galatolo2023right,acskin2023gendered,aljaroodi2023cultural} \\
\textbf{Cisgender} & Gender aligns with the gender originally assigned at birth/based on sex. & \cite{stolp2024morethanbinary,acskin2023gendered} \\
\textbf{Gender Nonconforming} & Modes of genderedness that buck norms and standards. & \cite{stolp2024morethanbinary,torre2023canagender,roesler2023hu} \\
\textbf{Transgender} & Gender does not align with the gender originally assigned at birth/based on sex. & \cite{stolp2024morethanbinary} \\
\textbf{Non-Binary} & Genders outside the binary; mediated by different socio-cultural contexts and times, e.g., third genders, pan-Indigenous Two-Spirit genders, X-gender (Japan). & \cite{huang2022proxemics,tolmeijer2021female,peck2020inducing,stolp2024morethanbinary,chang2024investigating,pitardi2023congruity,Telang2023,galatolo2023right,acskin2023gendered} \\
\textbf{[Agent] Gender} & Agent gender identity, potentially specific to the agent or agents in general and not humans. [Agent] can be replaced with specific agent type, e.g., robot, avatar. & \cite{otterbacher2017uncanny,reich2017irrelevance,dennler2023design,jung2016feminizing,weiss2020inconsequential,koda2017perception,peck2020inducing,stolp2024morethanbinary,robben2023theeffect,chang2024investigating,sandygulova2018agegender,perugia2022bias,ye2020effect,tay2014stereotypes,torre2023canagender,kuchenbrandt2012keepaneye,roesler2022context,choi2020avatars,deshmukh2018technology,pitardi2023congruity,koda2018perception,cameron2016congratulations,agren2022persuasiveness,shiomi2017audiovisual,thellman2018persuasive,wessler2021economic,bernotat2021female,Telang2023,parlangeli2023stereotypes,galatolo2023right,acskin2023gendered,sandygulova2015children,nomura2023expectations,tan2018isociobot,aljaroodi2023cultural,kervellec2016similarly} \\
\textbf{Gender Androgyny} & A concerted mixture of genders. & \cite{kraus2018effects,perugia2022bias,perugia2021gender,pitardi2023congruity,liu2023identity,galatolo2023right,dennler2023design}
\\
\textbf{Gender Ambiguity} & Gender-inconclusive, potentially mixed (feminine, masculine, others), and dependent on the individual. & \cite{seaborn2023can,tolmeijer2021female,kraus2018effects,perugia2022bias,Telang2023,liu2023identity,galatolo2023right,aljaroodi2023cultural} \\
\textbf{Gender Neutrality} & Lack of gender cues; gender-free by design. & \cite{jung2016feminizing,moreira2023can,rogers2020robotgendering,tolmeijer2021female,kuchenbrandt2012keepaneye,weiss2020inconsequential,seaborn2023can,robben2023theeffect,chang2024investigating,otterbacher2017uncanny,perugia2021gender,sandygulova2018agegender,roesler2023hu,parlangeli2023stereotypes,liu2023identity,acskin2023gendered,sandygulova2015children,nomura2023expectations,lee2021social,McGinn2020meet,tan2018isociobot,stroessner2019perception,dennler2023design} \\
\textbf{Gender Fluidity} & Gender is not fixed and can shift by time or context. & \cite{perugia2022bias,peck2020inducing,pitardi2023congruity} \\
\textbf{Agender/Non-Gendered} & Absence of a gender identity; genderless. & \cite{stolp2024morethanbinary,kraus2018effects,seaborn2023can,tolmeijer2021female,peck2020inducing,otterbacher2017uncanny,torre2023canagender,roesler2022context,perugia2021gender,pitardi2023congruity,acskin2023gendered,dennler2023design} \\
\bottomrule
\end{tabular}
\end{table*}

\begin{table*}[ht]
\caption{Gender and power.}
\label{tab:power}
\begin{tabular}{p{97pt}p{200pt}p{180pt}}
\toprule
Concept & Definition & Sources \\
\midrule
\textbf{Gender Stereotypes} & Simple, limited models of specific genders that percolate within socio-cultural settings (in line with the stereotype model by \citet{Bordalo2016stereo} using the subjective probability and representativeness heuristic theorizing of \citet{Kahneman1972}). & \cite{moreira2023can,peck2020inducing,chang2024investigating,kraus2018effects,torre2023canagender,parlangeli2023stereotypes,nomura2023expectations,galatolo2023right,Hwang2019sounds,otterbacher2017uncanny,reich2017irrelevance,tay2014stereotypes,choi2020avatars,acskin2023gendered,jung2016feminizing,weiss2020inconsequential,koda2017perception,seaborn2023can,tolmeijer2021female,stolp2024morethanbinary,robben2023theeffect,perugia2022bias,ye2020effect,roesler2022context,perugia2021gender,sandygulova2018agegender,roesler2023hu,agren2022persuasiveness,thellman2018persuasive,bernotat2021female,Telang2023,liu2023identity,kervellec2016similarly,stroessner2019perception} \\
\textbf{Gender Bias} & Preferential actions, attitudes, and perceptions for a certain gender, genders, or model of gender. & \cite{robben2023theeffect,jung2016feminizing,tolmeijer2021female,peck2020inducing,fitton2023dancing,chang2024investigating,sandygulova2018agegender,deshmukh2018technology,pitardi2023congruity,thellman2018persuasive,Telang2023,parlangeli2023stereotypes,galatolo2023right,nomura2023expectations,lee2021social,tan2018isociobot,dennler2023design} \\
\textbf{Gender Norms} & Norms (codified or informal) in roles and behaviours deemed appropriate for certain genders in certain cultural settings, societies, groups, and times (via \citet{butler2002gender} and \citet{Morgenroth2018}). & \cite{moreira2023can,kuchenbrandt2012keepaneye,Hwang2019sounds,stolp2024morethanbinary,robben2023theeffect,tay2014stereotypes,choi2020avatars,pitardi2023congruity,roesler2023hu,wessler2021economic,bernotat2021female,parlangeli2023stereotypes,liu2023identity,galatolo2023right,acskin2023gendered,aljaroodi2023cultural,dennler2023design} \\
\textbf{Gender Manipulation} & Influencing or controlling the act of gendering, e.g., via cues, in research or other contexts, with intent. & \cite{roesler2022context,jung2016feminizing,moreira2023can,chang2024investigating,reich2017irrelevance,sandygulova2018agegender,perugia2022bias,ye2020effect,tay2014stereotypes,kraus2018effects,kuchenbrandt2012keepaneye,perugia2021gender,pitardi2023congruity,agren2022persuasiveness,thellman2018persuasive,wessler2021economic,bernotat2021female,liu2023identity,galatolo2023right,sandygulova2015children,lee2021social,stroessner2019perception} \\
\bottomrule
\end{tabular}
\end{table*}

\begin{table*}[ht]
\caption{Methods to elicit gendering by design.}
\label{tab:design}
\begin{tabular}{lp{240pt}p{160pt}}
\toprule
Method & Definition & Sources \\
\midrule
\textbf{Gender Cue} & A salient feature of the agent's appearance (e.g., body or morphology, colour, clothing, anatomy), voice (e.g., speech pattern, pronouns), nonverbal behaviour, identity (e.g., name, social information), role, and personality that is linked to certain genders and elicits gendering (expanded from \citet{Tannenbaum2019}. & \cite{jung2016feminizing,otterbacher2017uncanny,perugia2022bias,thellman2018persuasive,torre2023canagender,choi2020avatars,koda2017perception,roesler2023hu,liu2023identity,sandygulova2018agegender,koda2018perception,tolmeijer2021female,Hwang2019sounds,moreira2023can,stolp2024morethanbinary,robben2023theeffect,tay2014stereotypes,kraus2018effects,roesler2022context,perugia2021gender,bernotat2021female,parlangeli2023stereotypes,galatolo2023right,acskin2023gendered,nomura2023expectations,tan2018isociobot,aljaroodi2023cultural,kervellec2016similarly} \\
\textbf{Gender Expression} & A holistic depiction or representation of gender in the design of the agent (in line with the Depiction Theory for robots~\cite{Clark2022depict}). & \cite{stolp2024morethanbinary,galatolo2023right,dennler2023design,tolmeijer2021female,moreira2023can,peck2020inducing,reich2017irrelevance,tay2014stereotypes,kraus2018effects,choi2020avatars,roesler2023hu,Telang2023,parlangeli2023stereotypes,liu2023identity,nomura2023expectations} \\
\textbf{Gender Traits} & A specific cue related to personality, nonverbal behaviours, and social interactivity, whereby certain modes of expression are linked to certain genders (as per the Gender Schema Theory of \citet{bem1981bem}). & \cite{perugia2022bias,kraus2018effects,tay2014stereotypes,koda2017perception,Hwang2019sounds,galatolo2023right} \\
\bottomrule
\end{tabular}
\end{table*}

\subsection{Manipulations} \label{result-agent-gender-manipulation}
Gender manipulation of the agent was present in 43 studies. We 
focused on four key dimensions: voice ($n=27$; \autoref{result-agent-voice}), body ($n=27$; \autoref{result-agent-body}), name ($n=12$; \autoref{result-agent-name}), and the gender options provided 
(\autoref{result-agent-genderOptions}). 

\subsubsection{Voice} \label{result-agent-voice}
Voice was used 
in 27 studies. 
Studies employed text-to-speech (TTS) systems, including commercial Application Programming Interfaces (APIs) like Google Cloud Text-To-Speech~\cite{huang2022proxemics}, Amazon Voices~\cite{weiss2020inconsequential, torre2023canagender}, and other TTSs~\cite{reich2017irrelevance, tay2014stereotypes, shiomi2017audiovisual, McGinn2020meet, kraus2018effects}. Specific TTS software like AITalk~\cite{koda2017perception, koda2018perception} and Acapela~\cite{sandygulova2018agerelated, kuchenbrandt2012keepaneye, sandygulova2018agegender, thellman2018persuasive, sandygulova2015children} were also used. Several studies featured computer-generated voices~\cite{rogers2020robotgendering, torre2023canagender, seaborn2023can}. Modulated voices were common, including modulated human voices~\cite{tolmeijer2021female, chang2024investigating} and robot voices~\cite{stolp2024morethanbinary}. Several used robot voices without further specification~\cite{robben2023theeffect, agren2022persuasiveness, liu2023identity}. Siri was used in one study~\cite{lee2021social}. Two studies mentioned ``female and male'' or ``female'' voices without clarifying whether these were synthesized, recorded, or robotic voices~\cite{perugia2021gender, tan2018isociobot}.

\subsubsection{Body} \label{result-agent-body}

Agent body was used 
in 27 studies, achieved through a variety of visual and morphological strategies that drew on culturally recognizable gender cues.
Several studies used fully human-like avatars or virtual agents to represent male or female identities~\cite{huang2022proxemics, koda2017perception, peck2020inducing, fitton2023dancing, koda2018perception, chang2024investigating, choi2020avatars, aljaroodi2023cultural}. Stylized or realistic human forms were used to align the agent's appearance with conventional gender norms.
Others used unmodified robot bodies, intended by default to be male, female, or neutral based on appearance alone~\cite{otterbacher2017uncanny, sandygulova2018agerelated, perugia2022bias, roesler2022context, parlangeli2023stereotypes, stolp2024morethanbinary}. Facial features were also used~\cite{stroessner2019perception, kervellec2016similarly, galatolo2023right, agren2022persuasiveness, perugia2021gender}.
Stereotypically gendered colours were used by some (e.g., pink for female, blue for male)~\cite{shiomi2017audiovisual, thellman2018persuasive, pitardi2023congruity, robben2023theeffect, jung2016feminizing}, while others intentionally used neutral tones~\cite{McGinn2020meet}.

More precise manipulations were used in a few studies. \citet{bernotat2021female} adjusted waist-to-hip ratios and shoulder widths to indicate gender. \citet{acskin2023gendered} designed robot bodies with broader shoulders for male agents, rounder shapes and smaller waist ratios for female agents, and intermediate proportions for neutrality.

\subsubsection{Name} \label{result-agent-name}
Only 12 studies explicitly used names to convey gender, typically drawing on cultural conventions. Several relied on binary name pairs: ``James'' and ``Mary''~\cite{rogers2020robotgendering}, ``Lucas'' and ``Luna''~\cite{robben2023theeffect}, ``Nero'' and ``Nera''~\cite{reich2017irrelevance, kuchenbrandt2012keepaneye}, ``John'' and ``Joan''~\cite{tay2014stereotypes}, and ``Alexander'' and ``Alexandra''~\cite{wessler2021economic}. In some cases, titles were added to reinforce gender cues, as with ``Mr. Robot'' and ``Mrs Robota''~\cite{pitardi2023congruity}.
A broader roster of culturally gendered names---``Fedora'', ``Arianne'', ``René'', ``Mei'', ``Anne'', ``Ursula'', ``Ted'', ``Fred'', ``Max'', ``August'', ``Marty'', ``Olaf'', and ``Geremy''---was used to probe gender perceptions based on name alone~\cite{perugia2021gender}.
A few used names to signal a specific \emph{or} ambiguous identity, as in ``Jen Brown''~\cite{choi2020avatars}, ``Stevie''~\cite{McGinn2020meet}, and ``Wynn''~\cite{stolp2024morethanbinary}. Flexible naming also appeared: ``Samantha'', ``Samuel'', and the non‑binary ``Sam'' were used in parallel~\cite{liu2023identity}, while the non-binary ``Sam'' was used exclusively in another study~\cite{torre2023canagender}.
Although less used than voice or body signals, naming was a meaningful strategy for signalling agent gender.

\subsubsection{Gender Options} \label{result-agent-genderOptions}
\begin{figure*}[!ht]
  \includegraphics[width=.55\textwidth]{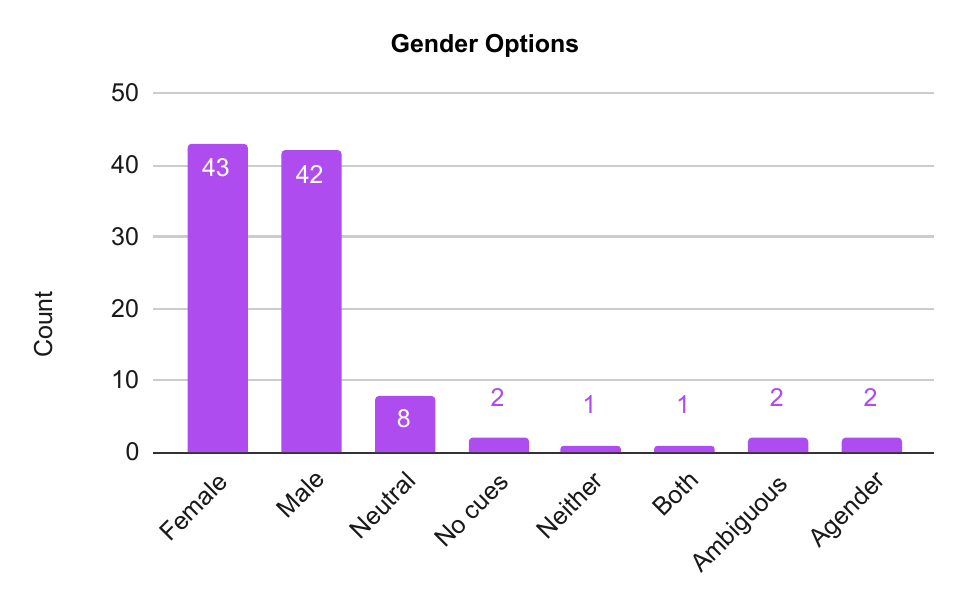}
  \caption{Agent gender options used by researchers.}
  \Description{Bar graph indicating that most relied on male or female options. Gender options included female, male, neutral, no cues, neither, both, ambiguous, and agender.}
  \label{fig:genopt}
\end{figure*}

Most studies included ``female'' (84.3\%) and ``male'' (82.2\%) categories (\autoref{fig:genopt}). Less commonly, options like ``neutral'' (15.7\%), ``ambiguous'' (two), ``agender'' (two), ``neither'' (one), and ``both'' (one) were also used.
Seven studies did not provide sufficient detail about how agent gender was manipulated.

\begin{figure*}[!ht]
  \includegraphics[width=\textwidth]{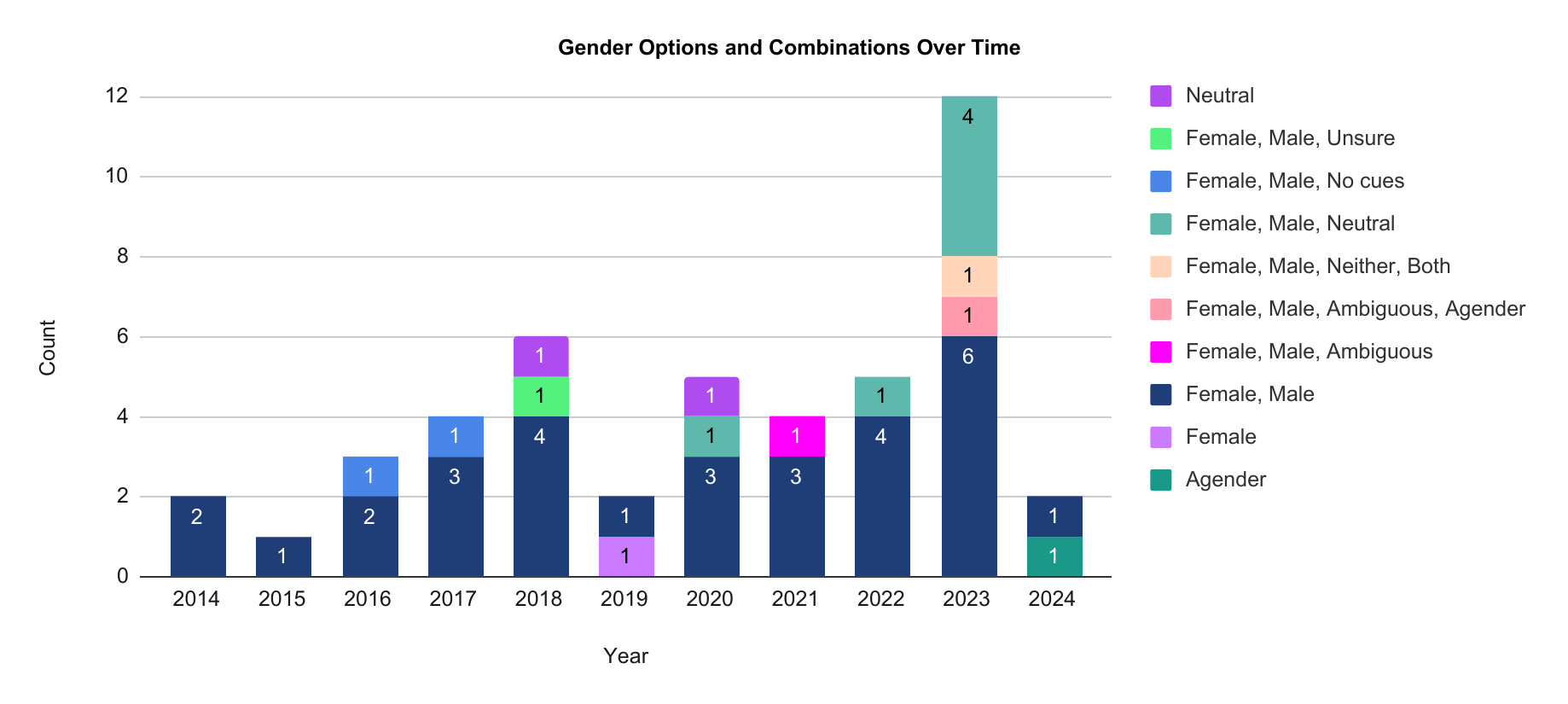}
  \caption{Agent gender options and combinations over time.}
  \Description{Stacked bar graph indicating that gender options and combinations of options appear to be diversifying over time.}
  \label{fig:genoptovertime}
\end{figure*}

Options seem to diversify over time (\autoref{fig:genoptovertime}). The first mention of a non-gendered term (outside ``female'' and ``male'') appeared in 2018~\cite{tan2018isociobot}, where ``neutral'' was used. In 2020, two studies adopted the term ``neutral''~\cite{rogers2020robotgendering, McGinn2020meet}, followed by a 2021 study using ``ambiguous''~\cite{tolmeijer2021female}. The term ``neutral'' was used again in 2022~\cite{perugia2021gender}. In 2023, several studies used a range of gender categories like ``agender''~\cite{torre2023canagender}, ``ambiguous''~\cite{torre2023canagender}, ``both'' and "neither"~\cite{seaborn2023can}, and ``neutral''~\cite{Telang2023, liu2023identity, acskin2023gendered, nomura2023expectations}. In 2024, one study used ``agender''~\cite{stolp2024morethanbinary}.

\subsection{Manipulation Checks}

\begin{figure*}[htbp]
    \centering
    \begin{subfigure}[b]{0.55\textwidth}
        \centering
        \includegraphics[width=\textwidth]{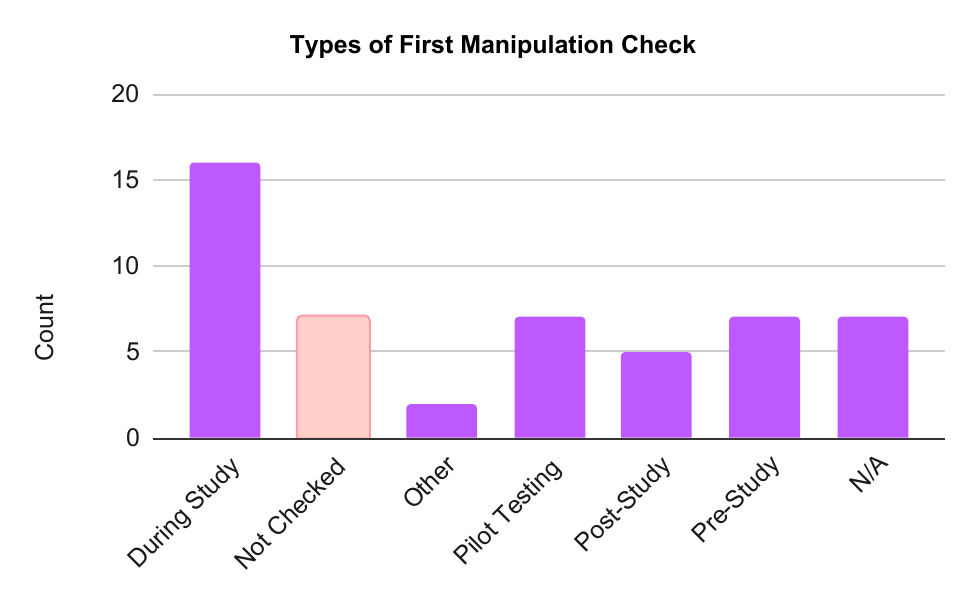} 
        \caption{First manipulation check.}
        \label{fig:mc1}
    \end{subfigure}
    \hfill 
    \begin{subfigure}[b]{0.312\textwidth}
        \centering
        \includegraphics[width=\textwidth]{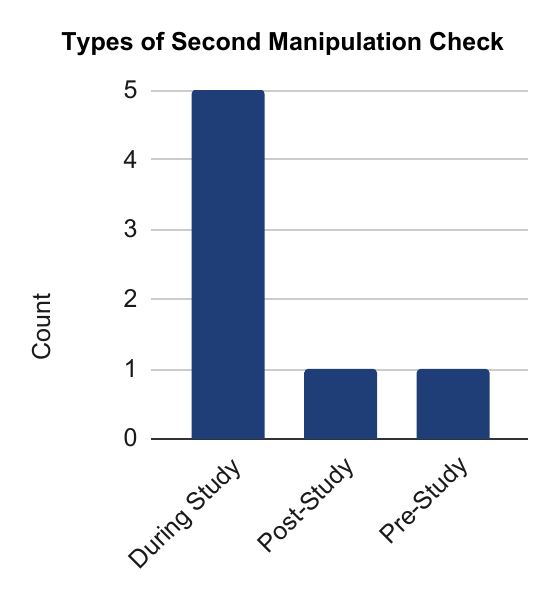} 
        \caption{Additional manipulation checks.}
        \label{fig:mc2}
    \end{subfigure}

    \caption{Manipulation checks, including counts of no checks.}
    \Description{Two figures indicating the type and spread of manipulation checks: during the study, not checked, other, pilot testing, post-study, pre-study, and N/A. The second figure indicates further manipulation checks within the same studies.}
    \label{fig:mc}
\end{figure*}

Forty-five manipulation checks were conducted across all studies (\autoref{fig:mc}). Seven papers did not conduct checks, seven did not report a check, seven reported a second check (\autoref{fig:mc2}), and one~\cite{kuchenbrandt2012keepaneye} reported three checks.

Type of manipulation check varied. Most used Likert-type items, with ratings via a feminine/masculine scale or similar ($N=20$) or checkboxes, i.e., male/female ($N=6$). One-offs included: naming the agent;
listing agent gender characteristics;
listing gender typicality of tasks;
a ``select the closest image that represents the agent'' item; and an IAT.
Twelve did not provide enough detail. Two used unrelated manipulation checks.

\subsection{Measures}
\label{sec:measures}
We report on how perceptions of agent gender were measured in 50 papers, excluding one~\cite{Hwang2019sounds} for use of a voice assistant defining itself. 
We report 56 measures (some used multiple measures). 
A summary is shown in \autoref{table:measureCats}.

\begin{table*}[!ht]
\caption{Summary of measures by purpose and type. Most used closed questions. Qual.: Qualitative. Quant.: Quantitative.}
\label{table:measureCats}
\begin{tabular}{l|rrrrr|r} \toprule
& \multicolumn{5}{c|}{Type of Measure} & \multicolumn{1}{l}{} \\
Purpose & Closed Question & Open Question & Unclear & Other Qual. & Other Quant. & Total \\ \midrule
Implicit Association & 0  & 0 & 0  & 0 & 1 & 1   \\
Gendered Language    & 0  & 0 & 0  & 4 & 0 & 3   \\
Similarity-to-Self   & 2  & 0 & 0  & 0 & 0 & 2   \\
Name & 0  & 2 & 0  & 0 & 0 & 2   \\
Gendered Traits & 23 & 0 & 0  & 0 & 0 & 23  \\
Assign Gender & 17 & 3 & 4 & 0 & 0 & 24 \\  
\midrule
Total    & 42 & 5 & 4  & 4 & 1 & 56  \\ 
\bottomrule
\end{tabular}
\end{table*}

\subsubsection{Implicit Associations}
One paper~\cite{roesler2022context} used the Implicit Association Test (IAT)~\cite{greenwald1998measuring}, which taps into unconscious associations and stereotypes rather than explicit opinions. \citet{roesler2022context} investigated associations between stereotypically masculine and feminine robots and occupations. 
The gendered stimulus of the agent must be associated with that gender by participants, and is unlikely to be useful in cases where gender is not a factor (e.g., associations with different non-binary identities are under-investigated).

\subsubsection{Gendered Language}
Four papers investigated the language used to refer to the agent. 
Three~\cite{weiss2020inconsequential, moreira2023can, liu2023identity} reviewed the language used without directly asking about gender, while the fourth~\cite{stolp2024morethanbinary} used an interview method to ask about gender and explore the responses. Two~\cite{weiss2020inconsequential, moreira2023can} used content analysis on open-ended responses to find trends in language use and the types of pronouns used for the agents. One~\cite{stolp2024morethanbinary} used a grounded theory approach to analyze the data, while another~\cite{liu2023identity} used gendered language coding schemes from prior research~\cite{boggs1999canary}. 
This type of analysis needs care, as gendered language is contextual and experts can be biased. Vetted coding schemes may alleviate bias.

\subsubsection{Similarity-to-Self}
Two papers measured how well an avatar matched or represented the participant, which 
we term ``similarity-to-self.'' 
One~\cite{fitton2023dancing} directly asked how closely the avatar's features matched the participant's own features using a Likert-type measure from 1 (least similar) to 5 (most similar). While 
indirect, similarity between a gendered avatar and the participant may be inferred.

The other~\cite{peck2020inducing} used the embodiment scale~\cite{peck2018effect}: four questions with two measuring embodiment and two agency via a 7-point Likert scale (1: strongly disagree, 7: strongly agree). The embodiment questions address the perception of an avatar's similarity-to-self via statements like ``I felt as if the body I saw in the virtual world might be my body.'' 
Gendering of the agent was again by inference.
However, the researchers treated embodiment and agency as a single measure, in contrast to the original work. This potentially led to the low Cronbach's $\alpha$ value ($-0.48$)~\cite{peck2020inducing}. A combined measure also reduces agent gender inferences, since agency only addresses the participant's control of the avatar.

Indirect similarity-to-self measures allow for inferences of agent gender relative to the participant's gender. 
However, this requires knowing how much the participant identifies with their own gender. The second example~\cite{peck2020inducing} did so using the Multi-Component Model of In-group Identification~\cite{leach2008group} and gender as the in-group. 

\subsubsection{Name}
Two papers~\cite{roesler2022context, roesler2023hu} tasked participants with providing a name to the agent using open-ended items. Both coded these names as male, female, nicknames (including typical animal names), and functional names. 
Thus, the extent to which an agent was gendered and how it could be inferred. However, the gender and cultural background of the coders can affect coding. Names, even ones with strong gender connotations, do not necessarily indicate a certain gender~\cite{Seaborn2023transce,Dev2021,Cao2021}. 
Also, non-gendered names may not be easily categorized, limiting assignations for non-binary genders~\cite{Cao2021}. As such, while allowing diverse answers that are not potentially skewed by asking about gender directly, this measure may be best used with others to support the recognition of gender and cultural diversity.

\subsubsection{Gendered Traits} \label{subsub:genderedtraits}
Perceptions of gendered traits---adjectives often associated with a certain gender---were measured in 23 papers. Some were explicitly coded, like ``masculine'' and ``feminine,'' while others were terms linked to gender in prior work, like ``agentic'' or ``warm.''
All papers used closed questions, but the type of question varied. This section gives an overview by type of question.

\noindent\paragraph{\textbf{Likert (Type) Scales}}
Twelve papers used one or more Likert (type) scales to assess the attribution of gendered traits. A Likert scale is a measure that asks for an item or statement to be rated from 1 (strongly disagree) to typically 5 or 7 (strongly agree). Terms other than agree/disagree can be used, e.g., ``not at all'' to ``frequently.'' 

Nine papers covered perceptions of masculinity and/or femininity. Six~\cite{chang2024investigating, tay2014stereotypes, galatolo2023right, dennler2023design, perugia2021gender} treated each as separate concepts in two scales. Plotted as two axes, neutral (low masculinity and low femininity) and both (high masculinity and high femininity) can be inferred. 
One~\cite{shiomi2017audiovisual} only considered femininity via the context of a robot's hug, assuming that a non-feminine hug would be masculine. Two~\cite{kraus2018effects, parlangeli2023stereotypes} seemed to use two Likert scales but were unclear.

Two papers considered masculine and feminine traits combined with other options. One~\cite{perugia2021gender} used three separate Likert scales: femininity, masculinity, and gender neutrality. Another ~\cite{torre2023canagender} used four Likert scales---masculine, feminine, agender, and ambiguous---with definitions 
provided
. Gender-expansive options~\cite{seaborn_exploring_2022} allow for comprehensiveness and perceptions of neutrality and ambiguity outside masculinity and femininity.

Two papers used Likert (type) measures to explore other gendered traits. One~\cite{otterbacher2017uncanny} had participants rate agent agency and experience. Agency (also called ``competence'') is a stereotypically masculine trait pertaining to the ability to achieve goals~\cite{fiske_universal_2007}. The second dimension taps into \textit{emotional} experience: warmth (or communal ability) is the stereotypically feminine trait of being personable, social, and sympathetic~\cite{fiske_universal_2007}. The second paper~\cite{chang2024investigating} measured perceptions of authority linked to masculinity. Such measures allow for making inferences without directly cuing the respondent to gender, potentially leading to more honest responses. However, stereotypes are culturally-situated and dynamic, and may fail for non-biased or unfamiliar participants.

\paragraph{\textbf{Semantic Differential}}
Seven papers used semantic differential scales to measure gendered traits perceptions, with two ends as opposing terms (e.g., small--large). 
Six papers used a single scale. Five~\cite{acskin2023gendered, agren2022persuasiveness, choi2020avatars, pitardi2023congruity, stroessner2019perception} treated masculinity and femininity as polar ends; however, this over-simplifies the concepts and prevents cross-gender or fluid measurements. 
One paper~\cite{acskin2023gendered} asked for two masculine--feminine ratings: the participant's personal view vs. society's view. Such a measure may reduce social desirability effects and yield more accurate perceptions~\cite{glick_ambivalent_2018}, although the difference between personal and society scores reflects the extent an individual has internalized social norms~\cite{acskin2023gendered}.
The sixth paper~\cite{koda2018perception} used one semantic differential scale for agent femininity (low--high), relying on the inference that a lack of femininity is equatable to masculinity.
The seventh paper~\cite{koda2017perception} used three scales to create a compound measure of gender (Cronbach's $\alpha=0.67$): masculine--feminine, hateful--lovable, and tough--delicate. These pairs assume that masculine and feminine are mutually-exclusive opposites.

\paragraph{\textbf{Multiple Choice}}
Two papers used multiple choice style questions.
One~\cite{seaborn2023can} asked participants to select one of the following: masculine, feminine, aspects of both, neither, or another option. ``Aspects of both'' was operationalized as gender ambiguity and ``neither'' as neutral (gender-free). Participants could write their own answer for the open-ended ``another option.'' Only one item could be selected.
Another paper~\cite{Telang2023} provided a list of gendered adjectives (e.g., friendly, strong, sassy). Participants were asked to select all that applied to the agent.
Both methods allow for variety in responses and avoided treating masculinity and femininity as mutually-exclusive terms. However, providing one option~\cite{seaborn2023can} risks participants believing they must give one answer.
Conversely, allowing multiple options~\cite{Telang2023} enables nuanced combinations of masculine and feminine perceptions. The lack of explicitly gendered terms also avoids alerting participants to the topic. Still, subjectivity enters in how the adjectives are assigned to different genders.

\paragraph{\textbf{Bem Sex Role Inventory}}
Three papers 
used items from the Bem Sex Role Inventory (BSRI)~\cite{bem1981bem}. The BSRI, developed in 1981, consists of three subscales (masculinity, femininity, and androgynous) with 20 items each (60 total). Likert scales are used to measure how often the subject displays a trait, with masculinity linked to agency and femininity linked to communality. All papers used different subsets of items from these subscales: nine from the short form in \citet{choi2009exploratory} (communion  Cronbach's $\alpha$ of $.777$, agency $\alpha$ of $.893$); the German version of the feminine/masculine subscales in \citet{eyssel2012s} (Cronbach's $\alpha$s of $0.91$ and $0.87$); and 14 traits based on the German~\cite{eyssel2012s} and another German~\cite{schneider1988erfassung} work in \cite{bernotat2021female} (Cronbach's $\alpha$s ranged from $0.861$ to $0.998$). Using different subscales renders future meta-synthesis work difficult. Instead, researchers should use the full BSRI~\cite{bem1981bem}, its short form~\cite{choi2009exploratory}, or a validated translation~\cite[e.g.,][]{schneider1988erfassung}
. However, the scale is outdated (at 45 years old) and the roles may lack contemporary relevance, so uptake of an alternative inventory may be ideal.



\subsubsection{Assign Gender}
Twenty-five measures asked participants to assign a gender using various methods.

Four papers~\cite{robben2023theeffect, ye2020effect, kervellec2016similarly, shiomi2017audiovisual} indicated a manipulation check or pre-test 
but were unclear.

Two measures involved open-ended questions. One paper~\cite{deshmukh2018technology} asked participants ``Did you think robot had a gender? If yes, what was the gender?'' While priming gendering, this method gave participants the option of ``no'' and was open-ended on the nature of the gendering rather than prescriptive. The other paper~\cite{lee2021social} used an open question to capture perceived gender in a manipulation check. The exact wording of the question was not specified.

Nineteen measures had closed questions.
Nine gave participants a set of gender options. One~\cite{wessler2021economic} only provided binary (male/female) options and used this as a manipulation check. Seven provided three options: (mostly) male and (mostly) female, but the third option varied. Three~\cite{sandygulova2018agegender, sandygulova2018agerelated, tolmeijer2021female}~\footnote{Note: \citet{sandygulova2018agegender} seems to be an extended version of \citet{sandygulova2018agerelated}.} had the option ``unsure,'' one~\cite{nomura2023expectations} had ``neutral,'' one~\cite{rogers2020robotgendering} had ``neither,'' one~\cite{huang2022proxemics} had ``non-binary/3\textsuperscript{rd} gender,'' and one~\cite{McGinn2020meet} had ``genderless.'' While closed questions aid consistency in answers across participants, they limit the ability to assign a gender beyond the researcher's pre-determined answers and thus may not fully capture all perceived genders.

Nine papers used Likert scales or semantic differential scales (refer to \autoref{subsub:genderedtraits}). One~\cite{aljaroodi2023cultural} used a Likert scale, where gender was assigned using two scales on whether the robot looked female/male (1: strongly disagree to 5: strongly agree). Six~\cite{thellman2018persuasive, tan2018isociobot, jung2016feminizing, reich2017irrelevance, kuchenbrandt2012keepaneye, roesler2023hu} used semantic differential scales, where one pole was (more) male and the other was (more) female. While most went from 1--5 or 1--7, one~\cite{roesler2023hu} went from 0--100. Another paper~\cite{cameron2016congratulations} used a semantic differential with the language ``A lot like a boy'' and ``A lot like a girl,'' as it was aimed at children. Children also had the option of ``not either.'' 
Such semantic differential scales can limit the ability to assign non-binary and multiple genders, although the inclusion of ``not either'' may somewhat overcome this limitation.

In the final two papers, participants selected the image that best represented the agent. One~\cite{sandygulova2015children} asked children to indicate whether they thought the agent was a man or woman. The other~\cite{liu2023identity} asked participants to choose from five images that would represent the agent and its gender: a human woman, human man, neutral human, cyborg, and cartoon robot. This method may be particularly beneficial when engaging populations with limited reading abilities, but may still cue 
the gender binary, since ``neutral'' human may be interpreted in different ways. 

\section{Discussion}

Gender is a social construct in our lived experiences with interactive agents, as well as in research. Our review of 10 years and 51 papers that approached agent gendering from a measurement standpoint also reveals how perceptions of agent gender can---or could be---operationalized: defined, theorized, and evaluated as a user percept. Considering agent gender operationalization as embedded in the process of research and design practice reflects its social structure~\cite{Risman2018gender,Risman2018structure}. The variety of models, domain standards, operationalizations defined by individuals or teams, and un/conscious biases about gender explain the diversity and general lack of standards so far. The quantitative reporting also reveals a disconnect between the scientific practice of developing standards, generalizing results, and validating constructs. We 
need to embark on a macro revolution in the research institution by interacting with each other consciously, reflecting on these disconnects, and developing common ground together. Doing so will align with principles of data feminism, whereby we consider how our practices are perpetuating inequalities and move towards using our work to challenge such power imbalances~\cite{dignazio2020data}. The synthesis we offer here is a first step towards that goal.


\subsection{On the Operationalization of Agent Gender Perceptions to Date}

Agent gendering is a clear topic of scientific and design interest in HAI practice.
A subset of this work is motivated by concerns around how negative behaviours and attitudes from the human world, e.g., gender stereotyping and toxicity~\cite{galatolo2023right,tay2014stereotypes,parlangeli2023stereotypes}, transfer to interactions with agents.
While we recognize the community's efforts to critically engage with this topic, our results also reveal gaps and a lack of common practice across researchers. This poses a significant challenge to basic knowledge synthesis and scientific rigour. In summary, we lack standardization in terms, theory, measures, and models. Specific issues identified in our results include:
\begin{itemize}
    \item a significant amount of work manipulated agent gender without measuring it, raising questions over the validity of any gender-based results (54 such papers were excluded from our review
    , i.e., almost half of quantitative agent gender-related research we identified is of questionable validity)
    \item the overwhelming majority of works failed to provide an explicit definition of gender as considered within the work, limiting readers' ability to contextualize results     
    \item exploration of agent gendering relies heavily on the masculine/feminine binary, unnecessarily limiting knowledge generation, raising questions of validity, i.e., have researchers captured the full breadth of user perceptions?, and artificially restricting the agent design space 
    \item little consistency in measures and methods for assessing agent gendering, despite clear overlaps, alongside disuse of existing validated measures; in total, 58 different measures across 51 papers
    \item inconsistent use of terms and sheer variety, which needs condensing and formalizing to create a shared vocabulary, distinct definitions, and clear ties to specific theories, which may be incongruous or in conflict
\end{itemize}

The salience of gender as a human identity and our power as designers and researchers to reify and/or challenge gender constructs---whether by enforcing a gender binary in our measures, or by portraying alternative ways of being ``masculine'' in the design of agent personas---means that agent gendering is, unavoidably, \emph{sociopolitical} and \emph{non-prescriptive}. This requires us to account for different worldviews 
so as not to preclude scientific rigour. Ultimately, we simply need to know each researcher's stance. We propose establishing common practice in how the community \textit{approaches} work on agent gendering at a meta-level: practices that acknowledge different conceptualizations of gender and support scientific rigour 
on the basis of methodological approach over (dis)agreement in worldview. Next, we offer one way to do this, grounded in the theoretical literature and findings from the surveyed work.

\subsection{Meta-Level Framework: Operationalizing Agent Gender as a Social Structure}
\label{sec:framework}

We can represent the gendering process of agents as a socially-grounded perceptual phenomenon and provide a standard high-level approach to operationalizing perceptions of agent gender in a meta-level framework comprised of (i) a unifying theoretical model and (ii) an abstract approach to operationalization.


\subsubsection{Theoretical Model}
\label{sec:model}

\begin{figure*}[!ht]
  \includegraphics[width=.5\textwidth]{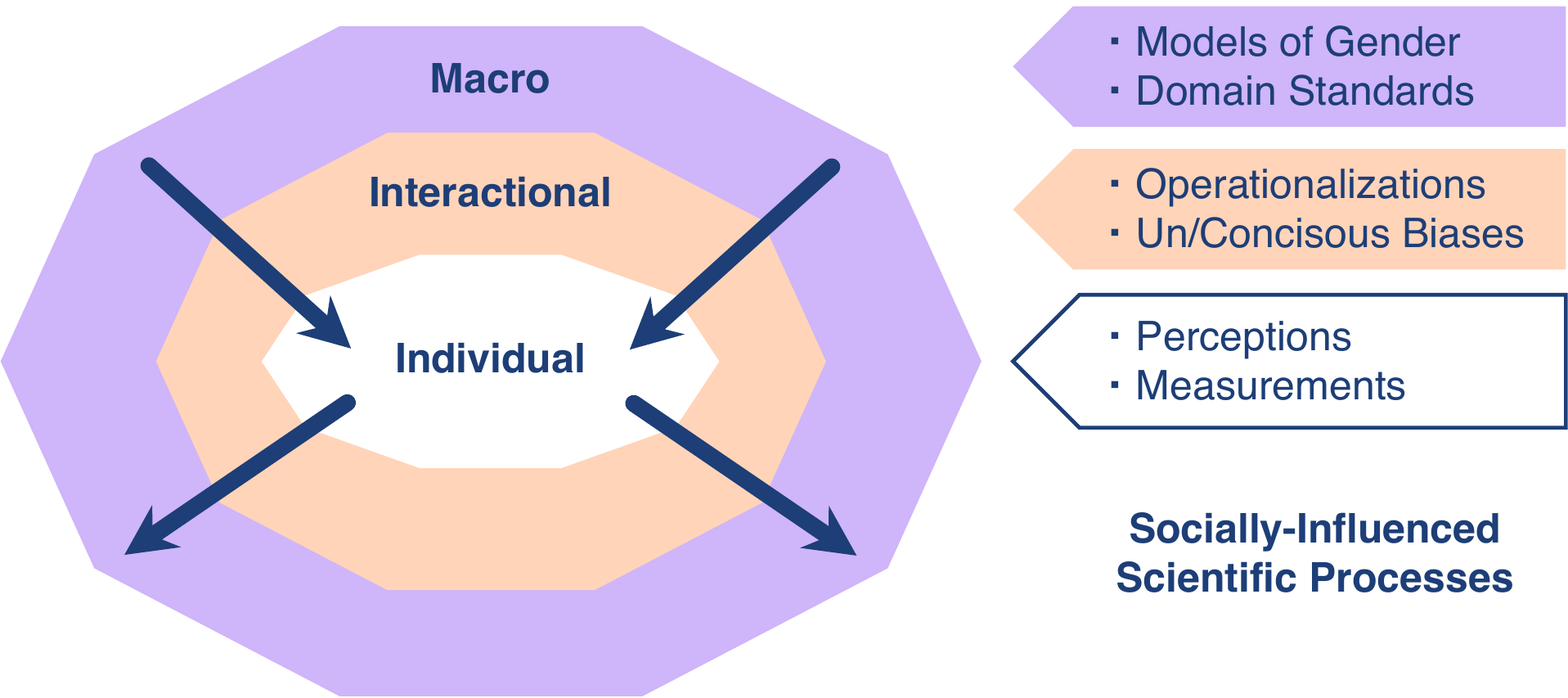}
  \caption{Agent gender as a social structure, extending the gender model by \citet{Risman2018gender,Risman2018structure} to the evaluation of HAI experiences in expert practice involving user perceptions measurement. Figure adapted from \citet{Steinhagen2025reconstruct} (with permission).} 
  \Description{The individual level is placed at the centre, inside of interactional, which is inside of macro. Arrows indicate that the individual influences and is influenced by the interactional and macro levels. Socially-influenced scientific processes include models of gender and domain standards at the macro level, operationalizations and un/conscious biases at the interactional level, and perceptions and measurements at the individual level.}
  \label{fig:structure}
\end{figure*}

\autoref{fig:structure} presents the theoretical model, adapted from the unifying theory of \emph{gender as a social structure} by \citet{Risman2018gender,Risman2018structure}. We have extended the basic model to include how science is influenced by the social processes involved in the structural model. Experts---roboticists, developers, designers, researchers, practitioners---as well as the targeted groups---end-users, participants, patients, co-designers---are influenced by macro-level models of gender, and experts in particular by domain standards. For instance, HCI researchers in feminist studies may have very different models of gender than other HCI researchers. Experts and non-experts will have their own operationalizations of gender---mental models and, for experts, possibly articulated technical definitions for research practice---that are implicated by un/conscious biases. This includes politically- and theologically-mediated decisions not to include or perceive certain gender options. Perceptions are measured at the individual level, even while they reflect the interactional context and the macro level---and potentially influence how others, including experts, understand gender.

Evaluating perceptions of agent gender, like all design work~\cite{CostanzaChock2018}, arises from an active and diverse community of practice. In kind, the model does not push for a particular term, definition, or theory of (agent) gender. Instead, it shows how
anyone operationalizing gender within a scientific and positivist framework needs to recognize and grapple with the realities of how gender works and the power experts wield as decision-makers of agent design and evaluation~\cite{Seaborn2023transce,CarstensenEgwuom2014,Seaborn2024botsagainstbias,winkle2023feminist}. We strongly advise taking on a design justice perspective when carrying out operationalization work, notably by considering the matters of expert power, intersectional disparities, macro-level influences, researcher bias, and marginalized perspectives and peoples~\cite{CostanzaChock2018}. In short, we on the expert side need to reflect on our power and centre the user, \emph{especially} when we feel certain that our worldview is the ``right'' one. Indeed, work within HAI has highlighted the need for greater care towards participants~\cite{schlesinger2017intersectional,winkle2021boosting} and also the machine~\cite{Seaborn2023transce}. This is a matter of ethics~\cite{winkle2023feminist,seaborn2025socialidentity,Seaborn2024botsagainstbias,spiel_patching_2019,Burtscher2020} \emph{and} scientific rigour~\cite{seaborn2025socialidentity,seaborn_unboxing_2025,seaborn_exploring_2022}. We must be transparent and inclusive of diversity, even if the user's take on gender differs from our own experiences, identities, and mental models of the world, not only to be moral agents but to do good research that truly captures reality as we \emph{all} know it.


\subsubsection{Approach to Operationalization}

We now offer a meta-level approach to operationalizing agent gender perceptions, one that can become standard practice and increase transparency, theoretical consistency, and methodological rigour. 
We situate our approach within the agent gender extension of the Risman model~\cite{Risman2018gender,Risman2018structure} introduced above (\autoref{sec:model}, \autoref{fig:structure}). 
The \textit{specific} models of (agent) gender available to the researcher and participants (\emph{macro}), the epistemological standpoint on gender and (un)conscious biases on both sides (\emph{interactional}), and each person's perceptions of agent gender (\emph{individual}) may vary by research question, local context, or life experience. However, we must recognize and make our \emph{conceptualization} processes and \emph{methodological} decision-making transparent to other researchers for understandability, comparability, and continuity in agent gendering work. 

As an example, gendered traits (c.f., ``agentic'' or ``warm'') was a common measure (24/50 papers). What traits used to signify agent gender for a specific participant (\emph{individual level}) relate to (un)conscious biases and conceptualizations of human gender in relation to agent gender (\emph{interactional level}), which in turn relates to mental models of gender as informed by, e.g., culture, education, and local legal definitions (\emph{macro level}). 
Importantly, the participant's worldview and personal experience may differ significantly from the researcher, who wields the power of defining, evaluating, and reporting.
Given the state of affairs 
and with the \emph{agent gender as a social structure} model as our guide, we advise rigorously and conscientiously approaching the operationalization of agent gender perceptions by reflecting on and transparently reporting
how gender has been considered within the context of the research. 

In practice, this requires \emph{four operationalization activities} relevant to the conceptualization and methodology stages (\autoref{fig:activities}). We derived these from standard practice when operationalizing phenomena of study~\cite{sage2017oper,sage2022oper}, modified for the case of agent gender as informed by our findings and the above model (\autoref{sec:model}). The four activities are:

\begin{figure}[!ht]
  \includegraphics[width=.475\textwidth]{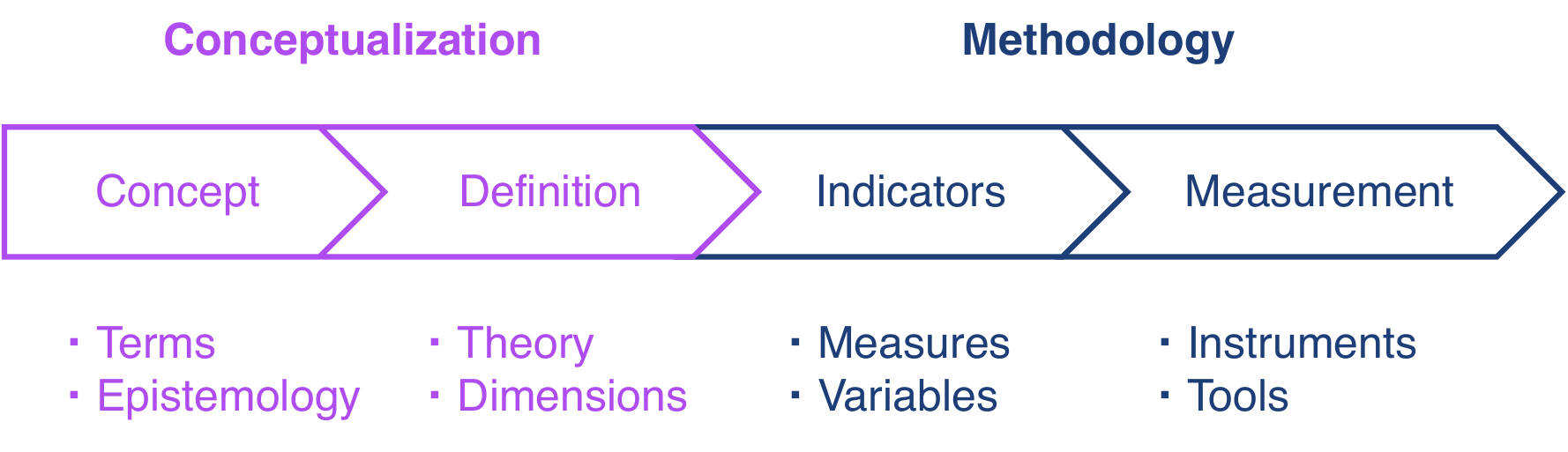}
  \caption{The four operationalization activities.}
  \Description{A four-part figure representing the proposed operationalization activities, comprised of conceptualization (concept and definition) and methodology (indicators and measurement).}
  \label{fig:activities}
\end{figure}

\begin{itemize}
    \item \textbf{Concept}: Provide the term/s and concept/s relevant to gender in the context of agent design or evaluation, linked to the aforementioned baseline definition of gender; we particularly stress the need for increased consistency in key terms (\autoref{tab:gendervocab} to \autoref{tab:design} provide a dictionary of terms to date) 
    \item \textbf{Definition}: Provide definition/s, referencing any appropriate theoretical or previous work
    \item \textbf{Indicators}: Outline the measures or variables used with justification and grounding in the conceptual dimensions outlined, e.g., the gender categories defined and relevance to the gender concept/s used (\emph{macro level}), potential biases of participants and researchers (\emph{interactional level}), and target perceptions 
    (\emph{individual level})
    \item \textbf{Measurement}: Detail how agent gender/ing was evaluated based on the indicators; if not the main objective, conduct and report manipulation checks to demonstrate that gender manipulations are received by participants as intended (if not in the paper, then in appendices, supplementary materials, or online repositories)
\end{itemize}

Evaluating perceptions of agent gender is a value-laden and political pursuit. The plurality in our surveyed dataset reflects this. Fifteen years ago, \citet{Bardzell2011methods} raised the challenge of balancing scientific extremes and traditions against the reality of sociocultural mediation and design subjectivity within HCI. The above approach enables internal consistency and demarcations among epistemological perspectives, again in recognition---as \citet{Bardzell2011methods} explicated---that epistemology informs choice of methods and thus frames any methodology.
Space is made for diversity in worldviews while supporting meta-analyses and knowledge synthesis. Notably, when reviewing the literature, researchers can identify works that do (or do not) match by epistemological standpoint.

A concrete example can be found in  \citet{acskin2023gendered}. They \textbf{conceptualize} gender based on the work of \citet{de1953second}, highlighting it as an act that people do. This framework is directly relevant to the study of how a robot's actions influence its gendering. They \textbf{define} gender as ``referring to masculinity and femininity and not as (assigned) sex''~\cite[p. 1917]{acskin2023gendered}. They give thorough consideration to gender schemas and stereotypes, grounding their hypotheses in relevant theories. Their \textbf{indicators} include two theory-derived measures of gender attribution---personal and societal---that cover the \emph{individual} and \emph{macro} levels while linking to the independent variable: the action being undertaken by the agent
. They also consider potential biases and assumptions at the \emph{interactional} level by pre-testing the stimuli they manipulated
. On \textbf{measurement}, they are consistent between their pre-tests and main study. Further, they report all details, including all scales and dimensions. The operationalization, from conceptualization to methodology, is clear, with all parts threaded together and woven into the overarching story of the paper. At a meta-level, it is solid.

\subsubsection{Developing New Tools for Evaluating Agent Gendering}
Our results identify an obvious need---and opportunity---for the development of new ways to evaluate agent gendering. First, 
we need to standardize and validate---for a particular conceptual definition of (agent) gender---the most common measures identified in \autoref{table:measureCats}, noting the limitations for each (\autoref{sec:measures}).
Going forwards, 
there is also ample room for the development of new methods of measurement. Drawing studies, which avoid the issue of prescription through a literal ``blank slate,'' may indirectly reveal agent gendering and allow for diverse representations~\cite{seaborn_unboxing_2025,decet2025sketch,seaborn_my_2025,lee_what_2019,straka2023sketchai} that can be purely qualitative or quantified through counts of features. Undertaking this approach would support the HCI as heterodoxy view that we should ``design so that we do as little as possible to hamper the evolution of variety''~\cite[p. 436]{light2011hci}. Still, we researchers must be reflexive and conscious of how our own worldviews may influence, e.g., the prompts to draw and interpretation of the drawings. A design justice perspective may guide a self-reflective stance that reminds us how every stage of the process is value-laden and challenges us to accept perceptions that we might not expect or agree with~\cite{CostanzaChock2018}. Development of any tools that aid in operationalizing perceptions of agent gender may ``encode''~\cite{CostanzaChock2018} our values and worldviews at the expense of others. We should aim for inclusion and equitability~\cite{CostanzaChock2018,schlesinger2017intersectional,winkle2023feminist}. When this is not possible, we must be transparent about it and provide solid justification. The meta-level framework provides a structure in which to do so.

\subsection{Limitations and Ethical Notes}
\label{sec:limitations}

We concluded our scoping review in the summer of 2024, so newer work was not included. Future reviews may resolve any differences or new trends. We divided the workload among us. Notably, only one researcher analyzed each part of the data solo. We recognize this as a major limitation because of the potential for unchecked bias during analysis. To control for bias, we used strategies like prompting the team for double-checks when done and being hyper-vigilant about our personal operationalizations of gender, resulting in regular requests for the team to question our interpretations. Still, we recognize that these strategies are not infallible, and there may have been missed opportunities without multiple coders. In consideration of reflexivity more broadly, we were limited by the English materials and Anglocentric, WEIRD frame, which commonly biases most research in HCI~\cite{Linxen2021weird}, HRI~\cite{Seaborn2023weird}, and the behavioural sciences~\cite{henrich_weirdest_2010}. We encourage explorations of similar frameworks outside of these frames. Additionally, we encourage work that directly engages intersectionality in agent gender-\emph{plus} perceptions~\cite{collins2020intersectionality,schlesinger2017intersectional,CostanzaChock2018}. Some work in our sample carried intersectional layers, like accent, age, and gender in Irish kids matched with gendered robots~\cite{sandygulova2015children} and the exploration of agent gender and emotional response, notably feminine anger, in \citet{wessler2021economic}. Still, perceptions of gender plus other social identities remain scant.

We stress that gender is a sensitive characteristic. Its measurement is subjective and political, bearing ethical questions~\cite{Burtscher2020,winkle2023feminist,bellini_feminist_2018}. While our focus was on perceiving gender in non-human agents, the constructivist nature of gender processing could reveal marginalized or politically-divergent mental models of gender in individuals. For example, non-binary and trans people may provide notably different gender perception ratings compared to cisgender people~\cite{Hope2022,stolp2024morethanbinary}. These ratings could, even when unlinked from participant gender, be reverse-applied to identify participant gender. Given the ongoing marginalization and violence worldwide against gender minorities, we emphatically recommend treating gender perceptions data as protected and highly sensitive.

\section{Conclusion}
We have systematically scoped out how user perceptions of agent gender have been operationalized. We have provided a map of terms, theory, and empirical studies. We have drafted up guidelines and proposed a meta-level framework that supports diverse modes of practice. The foundations are set. The next steps are fourfold. First, the community must take up this meta-level framework. Second, we must explore the options for standardizing specific operationalizations, of which we expect many based on the plurality in our sample. Third, we must come to consensus on these specific operationalizations. We must be open to future updates, since gender is a dynamic phenomenon in human societies and likely human--agent interaction work. 
Fourth, we must critique this praxis. The accuracy and vibrancy of the meta-level approach depends on the collective investment of everyone involved in HAI experiences.

If the meta-level framework is adopted, future meta-analyses can bring the ultimate goal of standard operationalizations to bear on the question of gender perceptions in HAI experiences. Only then will we know its efficacy and better understand, with confidence, how agent gender perceptions play out under specific operationalizations and epistemological frames. Likewise, our understanding of how people gender the world will be incomplete until then. We will also be unable to evaluate and take action on discrepancies and potentials, including those with impact: how agent gendering is implicated in behaviour and decision-making.

\begin{acks}
No acknowledgements or funding to declare.
\end{acks}

\bibliographystyle{ACM-Reference-Format}
\balance
\bibliography{REFS}

\newpage
\appendix

\onecolumn

\section*{Appendix}

\section{PRISMA-ScR Checklist}
\label{appendix:checklist}

\emph{Preferred Reporting Items for Systematic reviews and Meta-Analyses extension for
Scoping Reviews (PRISMA-ScR) Checklist from:}  Tricco A.C., Lillie E., Zarin W., O'Brien K.K., Colquhoun H., Levac D., et al. PRISMA Extension for Scoping Reviews
(PRISMAScR): Checklist and Explanation. Ann Intern Med. 2018;169:467–473. doi: 
\href{https://doi.org/10.7326/M18-0850}{10.7326/M18-0850}.


\begin{table}[!ht]
\caption{PRISMA 2020 Checklist.}
\label{table:prisma}
{\small
\begin{mytabular}[1.2]{|>{\raggedright}p{2.25cm}|>{\centering}p{0.5cm}|>{\raggedright}p{8.75cm}|p{2.15cm}|}

\hline
\rowcolor{purp}
\textcolor{white}{Section} & \textcolor{white}{Item} & \textcolor{white}{PRISMA-ScR Checklist Item} & \textcolor{white}{Reported on Page \#} \\
\hline 

\multicolumn{4}{l}{TITLE} \\
\hline
Title & 1 & Identify the report as a scoping review. & N/A; convention\\
\hline \hline

\multicolumn{4}{l}{ABSTRACT} \\
\hline
Abstract & 2 & Provide a structured summary that includes (as applicable): background, objectives, eligibility criteria, sources of evidence, charting methods, results, and conclusions that relate to the review questions and objectives. & N/A; convention\\
\hline \hline

\multicolumn{4}{l}{INTRODUCTION} \\
\hline
Rationale & 3 & Describe the rationale for the review in the context of what is already known. Explain why the review questions/objectives lend themselves to a scoping review approach. & \hl{pp. 1--2} \\ 
\hline
Objectives & 4 & Provide an explicit statement of the questions and objectives being addressed with reference to their key elements (e.g., population or participants, concepts, and context) or other relevant key elements used to conceptualize the review questions and/or objectives. & \hl{p. 2} \\
\hline \hline

\multicolumn{4}{l}{METHODS} \\
\hline
Protocol \& Registration & 5 & Indicate whether a review protocol exists; state if and
where it can be accessed (e.g., a Web address); and if available, provide registration information, including the
registration number & \hl{p. 4} \\
\hline
Eligibility criteria & 6 & Specify characteristics of the sources of evidence used
as eligibility criteria (e.g., years considered, language, and publication status), and provide a rationale. & \hl{p. 4} \\
\hline
Information sources & 7 & Describe all information sources in the search (e.g., databases with dates of coverage and contact with authors to identify additional sources), as well as the date the most recent search was executed. & \hl{p. 4} \\
\hline
Search & 8 & Present the full electronic search strategy for at least 1 database, including any limits used, such that it could be repeated. & \hl{p. 4} \\ 
\hline
Selection of sources of evidence & 9 & State the process for selecting sources of evidence (i.e., screening and eligibility) included in the scoping review. & \hl{p. 5} \\
\hline
Data charting process & 10 & Describe the methods of charting data from the included
sources of evidence (e.g., calibrated forms or forms that have been tested by the team before their use, and whether data charting was done independently or in duplicate) and any processes for obtaining and confirming data from investigators. & \hl{p. 5} \\
\hline
Data items & 11 & List and define all variables for which data were sought
and any assumptions and simplifications made. & \hl{p. 5} \\
\hline
Critical appraisal of individual sources of evidence & 12 & If done, provide a rationale for conducting a critical appraisal of included sources of evidence; describe the
methods used and how this information was used in any data synthesis (if appropriate). & \hl{N/A} \\
\hline
Synthesis of results & 13 & Describe the methods of handling and summarizing the
data that were charted. & \hl{p. 5} \\

\hline \hline
\end{mytabular}}
\end{table}

\begin{table}[!ht]
\caption{PRISMA 2020 Checklist, continued.}
\label{table:prisma2}
{\small
\begin{mytabular}[1.2]{|>{\raggedright}p{2.25cm}|>{\centering}p{0.5cm}|>{\raggedright}p{8.75cm}|p{2.15cm}|}

\hline
\rowcolor{purp}
\textcolor{white}{Section} & \textcolor{white}{Item} & \textcolor{white}{PRISMA-ScR Checklist Item} & \textcolor{white}{Reported on Page \#} \\
\hline 
\multicolumn{4}{l}{RESULTS} \\

\hline
Selection of sources of evidence & 14 & Give numbers of sources of evidence screened, assessed for eligibility, and included in the review, with reasons for exclusions at each stage, ideally using a flow diagram. & \hl{p. 5 and Fig. 1} \\ 
\hline
Characteristics of sources of evidence & 15 & For each source of evidence, present characteristics for which data were charted and provide the citations. & \hl{N/A; not evidence-based} \\ 
\hline
Critical appraisal within sources of evidence & 16 & If done, present data on critical appraisal of included sources of evidence (see item 12). & \hl{N/A} \\ 
\hline
Results of individual sources of evidence & 17 & For each included source of evidence, present the relevant data that were charted that relate to the review questions and objectives. & \hl{pp. 5--13} \\ 
\hline
Synthesis of results & 18 & Summarize and/or present the charting results as they relate to the review questions and objectives. & \hl{pp. 5--13} \\ 

\hline 
\multicolumn{4}{l}{DISCUSSION} \\
\hline
Summary of evidence & 19 & Summarize the main results (including an overview of concepts, themes, and types of evidence available), link
to the review questions and objectives, and consider the relevance to key groups. & \hl{pp. 13--15} \\ 
\hline
Limitations & 20 & Discuss the limitations of the scoping review process. & \hl{p. 15} \\ 
\hline
Conclusions & 21 & Provide a general interpretation of the results with respect to the review questions and objectives, as well as potential implications and/or next steps. & \hl{p. 16} \\ 
\hline 

\multicolumn{4}{l}{FUNDING} \\
\hline
Funding & 22 & Describe sources of funding for the included sources of evidence, as well as sources of funding for the scoping review. Describe the role of the funders of the scoping review. & \hl{p. 16 (N/A)} \\ 

\hline \hline
\end{mytabular}}
\end{table}

\end{document}